# Ferrocene as an iconic redox marker: from solution chemistry to molecular electronic devices

Gargee Roy[1,#], Ritu Gupta[2,#], Satya Ranjan Sahoo[3,4], Sumit Saha[3,4], Deepak Asthana[1*], Prakash Chandra Mondal[2*]

[1]Department of Chemistry, Ashoka University, Sonipat, Haryana 131029, India

[2]Department of Chemistry, Indian Institute of Technology Kanpur, Uttar Pradesh 208016, India

[3]Materials Chemistry Department, CSIR-Institute of Minerals & Materials Technology, Bhubaneswar, Odisha 751013, India

[4]Academy of Scientific and Innovative Research (AcSIR), Ghaziabad, Uttar Pradesh 201002, India

[#] These two authors equally contributed to this work

Corresponding author's E-mail: pcmondal@iitk.ac.in (PCM), deepak.asthana@ashoka.edu.in (DA)

**Abstract:** Ferrocene, since its discovery in 1951, has been extensively exploited as a redox probe in a variety of processes ranging from solution chemistry, medicinal chemistry, supramolecular chemistry, surface chemistry to solid-state molecular electronic and spintronic circuit elements to unravel electrochemical charge-transfer dynamics. Ferrocene represents an extremely chemically and thermally stable, and highly reproducible redox probe that undergoes reversible one-electron oxidation and reduction occurring at the interfaces of electrode/ferrocene solution in response to applied anodic and cathodic potentials, respectively. It has been almost 70 years after its discovery and has become one of the most widely studied and model organometallic compounds not only for probing electrochemical charge-transfer process but also as molecular building blocks for the synthesis of chiral organometallic catalysts, potential drug candidates, polymeric compounds, electrochemical sensors, to name a few. Ferrocene and its derivatives have been a breakthrough in many aspects due to its versatile reactivity, fascinating chemical structures, unconventional metal-ligand coordination, and the magic number of electrons (18 e$^-$). The present review discusses the recent progress made towards ferrocene-containing molecular systems exploited for redox reactions, surface attachment, spin-dependent electrochemical process to probe spin polarization, photo-electrochemistry, and integration into prototype molecular electronic devices. Overall, the present reviews demonstrate a piece of collective information about the recent advancements made towards the ferrocene and its derivatives that have been utilized as iconic redox markers.

# 1. Introduction: eccentric chemistry of ferrocene for the chemists

Fisher and Wilkinson received the Nobel prize in Chemistry for their pioneering work on 'sandwich organometallic compounds' back in 1973.[1] However, the journey of ferrocene chemistry was started independently by Kealy and Pauson and accidentally discovered ferrocene in 1951.[2,3] They attempted to synthesize 'fulvalene' via oxidative coupling reaction between cyclopentadienyl magnesium bromide and ferric chloride ($FeCl_3$) in anhydrous ether. Instead, the reaction produced a thermally stable (up to 175°C), an orange-colored sandwiched compound which is known as ferrocene ($C_{10}H_{10}Fe$, Fc). However, at that time chemical structure and metal-ligand boning in ferrocene were little-known. The correct structure was first proposed by none other than G. Wilkinson in 1952.[4] The research team led by Wilkinson suggested that $sp^2$-hybridized carbon atoms in cyclopentadienyl anion ring might be capable of forming a symmetrical covalent bond with the d orbitals of $Fe^{2+}$ resulting in a sandwich compound structure which was a great surprise to the chemistry community. Along with other analytical pieces of evidence, the infrared spectroscopy indicates the presence of only one type of C-H bond appears at 3050 cm$^{-1}$. The sandwich structure was further confirmed by several other research groups and since then, ferrocene and its structure received huge attention.[5–8] Depending on the potential energy, ferrocene can exist either lower energy staggered conformation (S-Fc, $D_{5d}$) or high energy eclipsed conformation (E-Fc, $D_{5h}$). **Fig. 1** and **2** represent an orbital charge density and molecular orbital (MO) diagram and details discussions can be found in the literature.[9,10] From **Fig.1**, the metal-ligand (M-L, Fe-C) covalent bonding features can be observed in the ferrocene, and p-orbitals of cyclopentadienyl (Cp) form unconventional π-bonded interactions. In the Cp ring, the electron density of all five carbon atoms is delocalized. Ferrocene being reversibly redox-active which undergoes Fc to $Fc^+$ (via one electron oxidation), rich coordination chemistry, and non-linear optical response has been extensively used to prepare stimuli-responsive molecular systems for various applications.[11–15] Interestingly, ferrocene also allows the ligand exchanges in which one of the Cp rings could be replaced with aromatic.[16] Ferrocene functions as an excellent bridge in setting electronic communications between two distant groups and used in ferrocene-peptide chemistry for investigating redox process.[17,18]

Ferrocene exhibits rich electrochemical behaviour that includes stable, and reversible Faradaic process, fast charge-transfer kinetic, oxidation to reduction current density is nearly unity ($I_{pa}/I_{pc} \approx 1$), peak-to-peak difference, $\Delta E_p$ is very close to ideal value (59 mV for 1e$^-$ transferred redox-process), well-defined half-wave redox signal, $E_{1/2}$ at +0.31 V (SCE), similar diffusion coefficient for both oxidized and reduced species.[19] First polarographic behaviour of metallocenes including ferrocene/ferrocenium was reported with respect to the saturated calomel electrode (SCE) as the reference electrode.[20] However, SCE creates liquid junction potential (LJP) between the electrode surface and the aqueous solvent which differ from solvent to solvent and geometry changes of the reference electrode. To avoid such LPJ formation, there was a pressing need for an internal redox standard which can provide reliable electrochemical redox process and the potential of the other redox-active systems can be expressed with respect to such internal redox marker. Considering its well-known, distinct electrochemical process discussed above, ferrocene has been employed as an internal redox-probes for calibrating potentiostat, reference electrodes, electrochemical stability of many redox-active species.[21–24] Besides, ferrocene has been used as a known electrochemical reference to deduce the energy levels of the highest occupied molecular orbital (HOMO) and lowest unoccupied molecular orbital (LUMO) of either organic, organometallic, or coordination compounds.[25,26] Ferrocene has not only been used as standard redox-probe, but also used as blocking agent to quantify surface coverage, and quality of the molecular thin films grafted on different electrodes.[27] In the class of stimuli-responsive systems, redox-active materials have attracted huge attention as the reversible oxidation/reduction process leads to alteration in electronic properties of the system and therefore extremely useful in devices in which the functionalities depend on electron acceptance/donation. In particular, optical changes in response to the redox activity make such systems ideal for electrochromic and optoelectronic applications. Thus, ferrocene is among one of the most attractive redox-markers that find diverse applications. In the present review, we will discuss the progress seen in the last five years mad with the ferrocene and its derivatives covering some recent aspects schematically shown in **Fig. 3**. This review would facilitate novice and expert to have in-depth understanding of electrochemical process and recent advanced achieved with ferrocene which is considered as an iconic redox marker.

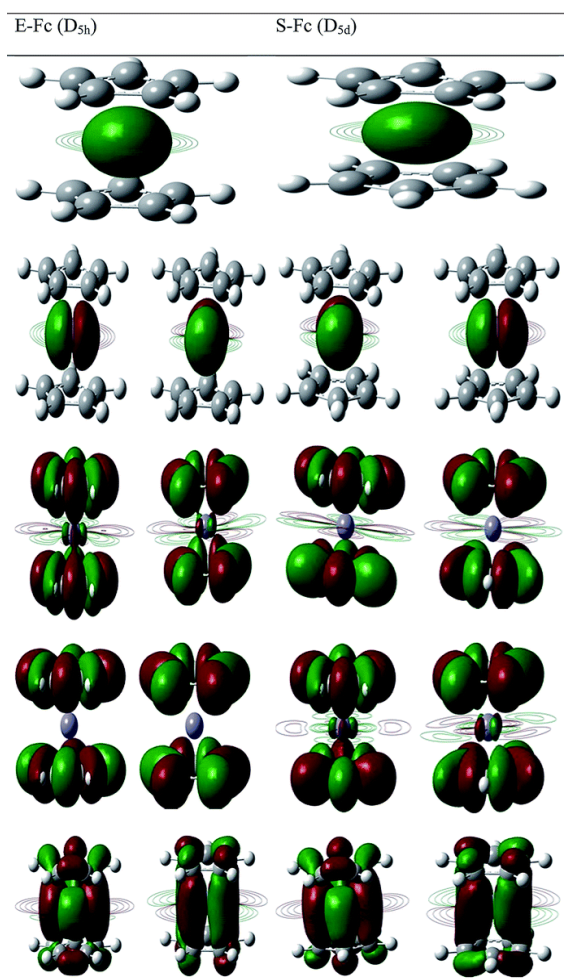

**Fig. 1.** DFT calculated orbital charge densities of Fc in two different configurations, $D_{5h}$ and $D_{5d}$. Reproduced with permission from Ref. [9]. Copyright 2015, Royal Society of Chemistry.

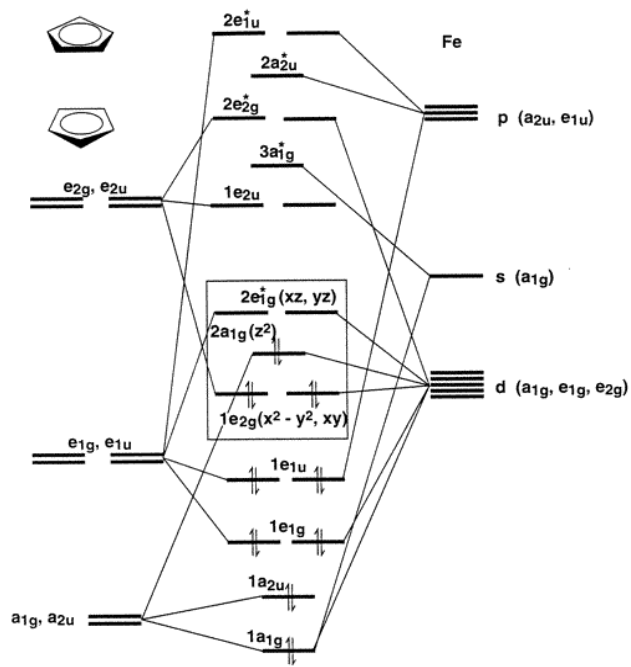

**Fig. 2.** Molecular orbital diagram of ferrocene. A total no. of 18 electrons is originated from $Fe^{2+}$ (6 e$^-$) and cyclopentadienyl ligands (6 e$^-$ from each). Reproduced with permission from Ref. [10]. Copyright 1999, American Chemical Society.

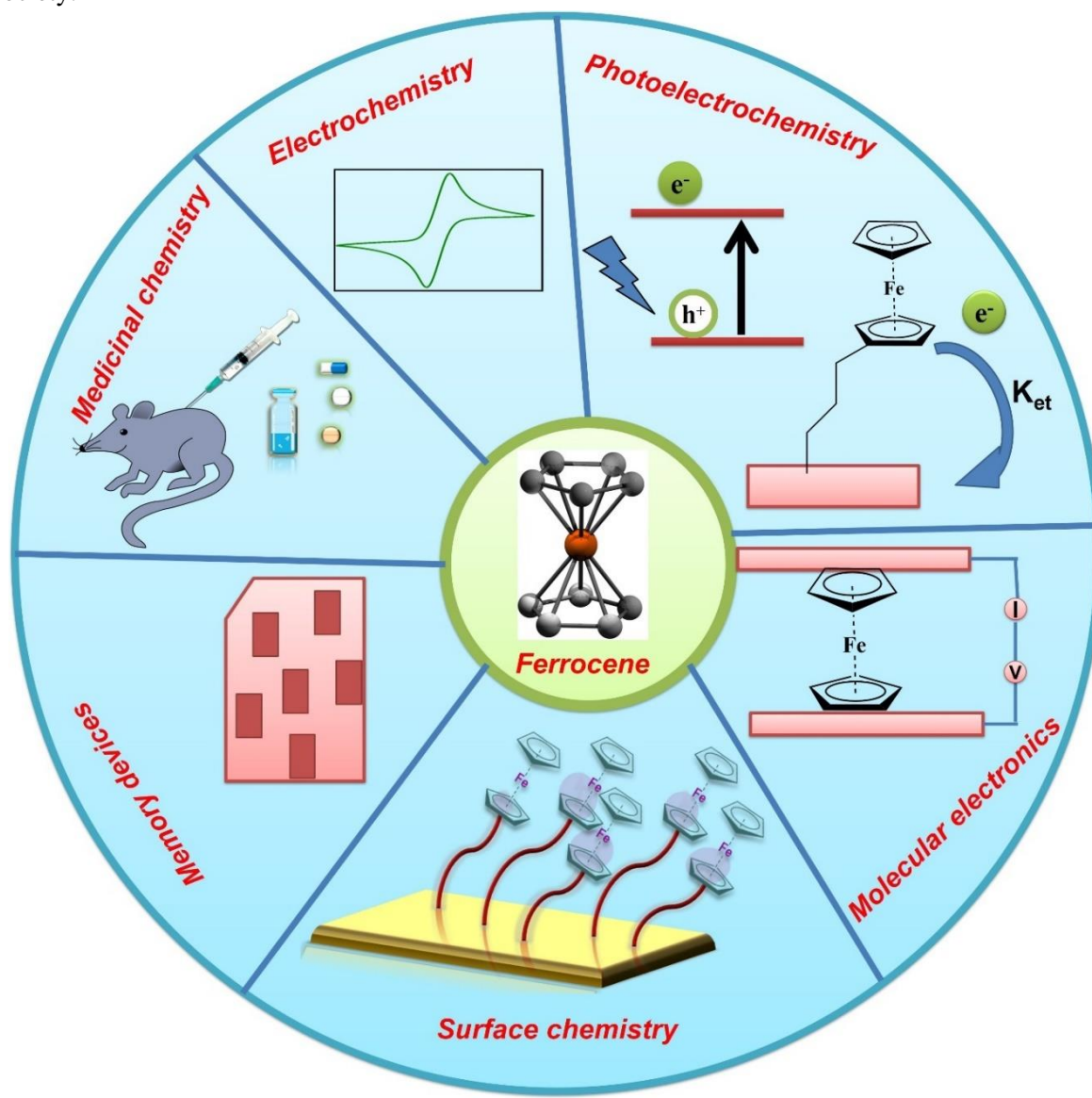

**Fig. 3.** Selective but recent domains achieved with ferrocene. Few of them are the key subject of the present Reviews.

## 2. Redox behavior of ferrocene: roles of working electrodes, Fermi energy, and charge-transfer mechanisms

In cyclic voltammetry (CV), we typically use a three-electrode set-up, working electrode (WE), reference electrode (RE), and counter electrode (CE). Working electrode carries all redox processes of our interests, mainly occurs at the electrode-electrolyte interface. Whereas reference electrode is used as a reference point with respect to we measure applied potential to the working electrode and counter electrode completes the circuit, i.e., current flows from working to counter electrode.[28] In all electrochemical studies, we are mainly concerned with all redox events and factors that affect these events. There are various factors that affect the redox process, such as types of working electrodes, Fermi energy ($E_F$), electrolytes and concentration, temperature and charge transport mechanism.[29] Every working electrode has a characteristic $E_F$ value; as a result, it has a different energy barrier for charge-transfer to happen. Now, consider an electrochemical setup where working, reference, and counter electrodes are immersed in an electrolyte solution containing ferrocene as a redox probe; when a potential difference is applied between working and reference electrodes, current flows in the external circuit due to the chemical reactions (redox) of ferrocene at electrode/electrolyte interface.[30] **Fig. 4** depicts the energy profile diagram of redox events of ferrocene. In response to an applied positive potential (+V vs RE), $E_F$ of the working electrode is modulated to nearer to the HOMO of ferrocene, as a result there will be an energetically favourable situation and ferrocene will transfer one d-electron to the working electrode ($d_z^2$ to $E_F$) forming ferrocenium, Fc (**Fig. 4a**). Similarly, the $E_F$ of the electrode can be enhanced by applying a negative potential (-V) and the releases electron back to the Fc, thus a reduction process happens. However, if the applied potential is too much negative, the ligand (Cp ring) gets reduced further. For the sake of simplicity, we would consider only $Fe^{2+/3+}$ electrochemical process but not the ligand-based electrochemical process.

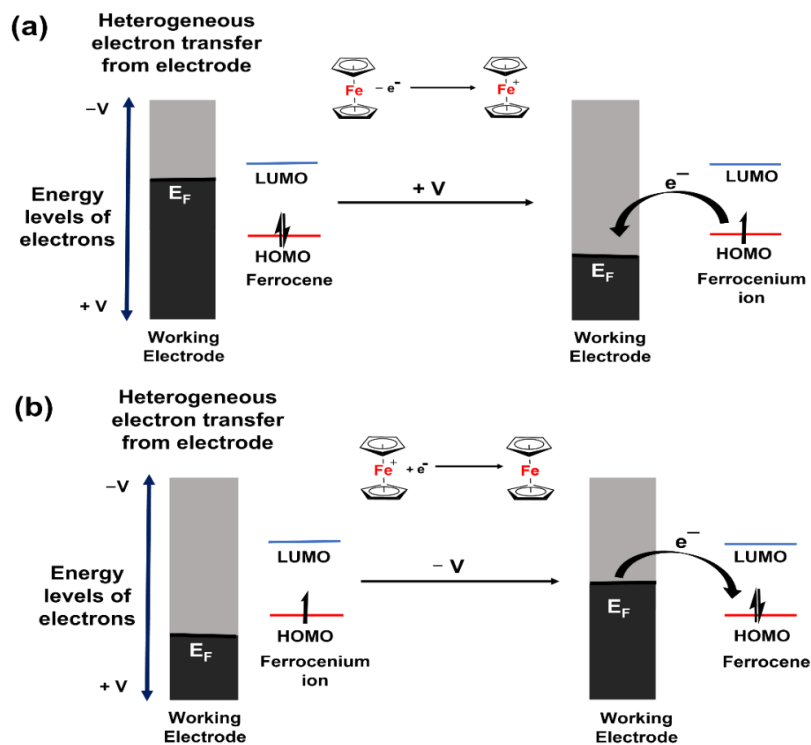

**Fig. 4.** Schematic representation of energy profile diagram of redox events of ferrocene at working electrode/electrolyte interface.

## 3. The solution-based electrochemical process in ferrocenes

Ferrocene can easily be functionalized on one or both Cp rings with the same or different substituents. Such functionalization does not affect electrochemical reversibility, or structural motif but this does affect its redox potential. For instance, substituting ferrocene with an electron-withdrawing group increases the oxidation potential, whereas electron-donating group decreases.[31,32] Several studies aim to prepare ferrocene derivatives for studying their electrochemical behavior and its discrepancy from the pristine ferrocene.[33] Bennett *et al.* synthesized a series of conjugated alkenyl arene ferrocene and biferrocene derivatives using four different ligands (**Fig. 5a**), namely, 4-ethynylpyridine (a), 3-ethynylpyridine (b); 4-ethynylthioanisole (c) and 3-ethynylthioanisole (d), and systematically investigated their electrochemical property by cyclic voltammetry (CV) measurements.[34] The CV experiments were conducted in dichloromethane solutions using 0.1 M N$^n$Bu$_4$PF$_6$ as the supporting electrolyte. The CV results revealed that monoferrocene compounds showed the characteristic one-electron redox process like ferrocene, but the oxidation potential was shifted to more positive side than the standard ferrocene (**Table 1** and **Fig. 5b,c**). It was attributed to the electron-withdrawing effect of ethynylarene substituents that withdraws the electron density from the ferrocene system. This effect can be clearly seen in **1cc** and **2c**. Furthermore, the substituents in arene also affect the potential and the reversibility of the redox process by stabilizing the positive charge through the donation of electrons into the π-system. Both pyridyl- and thioanisole-based analogs showed a similar but opposite effect. It was observed that delocalization of positive charge was effective in the case of 4-thioanisole as compared to 3-thioanisole, which clearly resembles with redox potential trends (**1dd** and **1cc**), showing higher potential for 3-thioanisole than 4-thioanisole. Similarly, 3-pyridyl showed more effective delocalization than that of 4-pyridyl, and the analog containing both groups (**1ab**) showed intermediate potential.

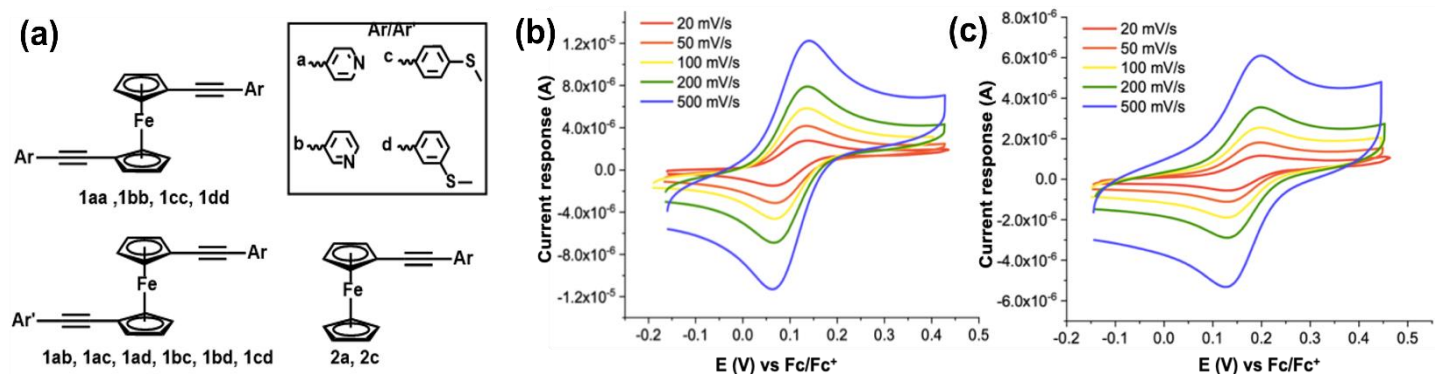

**Fig. 5.** (a) Different synthesized ferrocene derivatives used for electrochemical studies. (b) CVs of **1cc** and (c) **2c** were recorded in $CH_2Cl_2$/0.1 M [$nBu_4$][$PF_6$] recorded on a glassy carbon disc as the working electrode, Pt-wire as reference electrode. Reproduced with permission from Ref. [34]. Copyright 2021, American Chemical Society.

**Table 1**. Electrochemical parameters of **1** and **2** recorded at 100 mV s$^{-1}$ in 0.1 M $^n$Bu$_4$NPF$_6$ in $CH_2Cl_2$ (all potentials are in mV units). Reproduced with permission from Ref. [34]. Copyright 2022, American Chemical Society.

| [Fc] | $E_p$ | $E_c$ | $E_{1/2}$ | $\Delta E$ | $i_{pa}/i_{pc}$ |
|---|---|---|---|---|---|
| **1aa** | 336 | 264 | 300 | 72 | 1.23 |
| **1ab** | 319 | 249 | 284 | 70 | 1.06 |
| **1ac** | 247 | 182 | 215 | 65 | 1.08 |
| **1ad** | 284 | 213 | 249 | 71 | 1.11 |
| **1bb** | 296 | 209 | 253 | 87 | 1.05 |
| **1bc** | 246 | 172 | 209 | 74 | 1.02 |
| **1bd** | 268 | 188 | 228 | 80 | 0.89 |
| **1cc** | 196 | 130 | 163 | 66 | 0.93 |
| **1cd** | 196 | 130 | 163 | 66 | 0.86 |
| **1dd** | 237 | 173 | 205 | 64 | 1.00 |
| **2a** | 195 | 127 | 161 | 68 | 0.98 |
| **2c** | 135 | 69 | 102 | 66 | 0.94 |

A similar kind of studies were performed by Sauvage and coworker where they have synthesized a series of ferrocene-based compounds and their corresponding oxidized derivatives and evaluated their electrochemical properties by cyclic voltammetry (**Fig. 6**).[35] The CV plots revealed that by introducing simple electron donating groups like methyl or tert-butyl units on the Cp-rings, the redox potential was lowered as compared to pristine ferrocene. With the increase in number of methyl groups on Cp-rings, the redox potential was gradually decreased by ca. 100 mV (one methyl group per one Cp ring), that is, +0.302 V (vs SCE) for $Me_2Fc^+/Me_2Fc$ and +0.003 V for $Me_8Fc^+/Me_8Fc$. Therefore, the $Fe^{+III}/Fe^{+II}$ redox potentials were tuned from +0.403 V (ferrocene) down to −0.096 V ($Me_{10}Fc^+/Me_{10}Fc$) vs SCE.

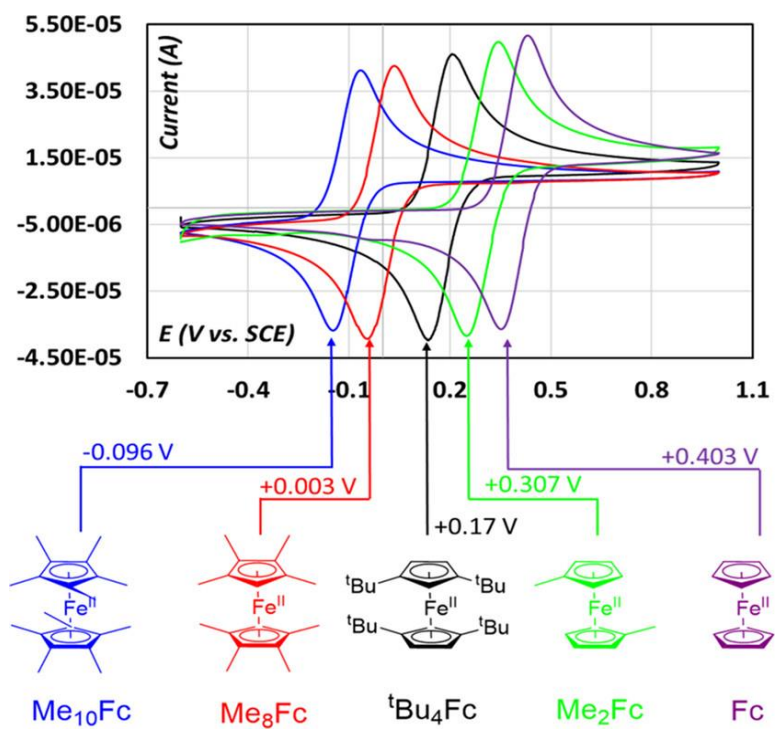

**Fig. 6.** Cyclic voltammogram and redox potential of Fe(III)/Fe(II) redox couple of ferrocene derivatives. Reproduced with permission from Ref. [35] Copyright 2019, American Chemical Society.

Monomeric zinc ferrocene carboxylate (Zn(FcCOO)(3,5-dmp)$_2$Cl) complex also exhibited one-electron redox reaction with a half-potential, E$_{1/2}$ at 442.5 mV, which was shifted towards more negative as compared to ferrocene, due to the electron-withdrawing nature of the carboxylate and the coordination bond between the carboxylate and Zn$^{2+}$ ions.[36] The value of $\Delta E_p$ was found to be 183 mV which indicated an one-electron reversible process. Moreover, a unique star-shaped ferrocene derivative, namely tetra[4-(4-ferrocenyl-1,2,3-triazol-1-yl)butyl]ferrocene was reported, which displayed the electrochemical reversibility with $\Delta E_p \leq 0.08$ V (= E$_{pa}$-E$_{pc}$).[37] Increasing the number of Fc units in these derivatives had an increased effect on both oxidation and reduction potentials and Faradaic currents. Wang *et al.* functionalized BODIPY (**3**, **4**, and **5**) with bisferrocenyl ligand to study the impact of ferrocenyl substitution on the electronic properties of BODIPY (**Fig. 7**).[38] **3** and **4** exhibited two distinguishably reversible one-electron oxidation peaks corresponding to two ferrocene units in these complexes. The oxidation potential was shifted towards more negative side as compared to the Fc/Fc$^+$ redox couple due to σ-donation from boron unit. **5** exhibited one two-electron oxidation peak for two ferrocene units in the complex, which was attributed to the decreased electronic communication or electrostatic effect owing to the long distance between two ferrocene units resulting in from the alkyne bridge. The oxidation potential was shifted towards positive as compared to the ferrocene/ferrocenium couple due to the electron-withdrawing effect of the alkyne bridge.

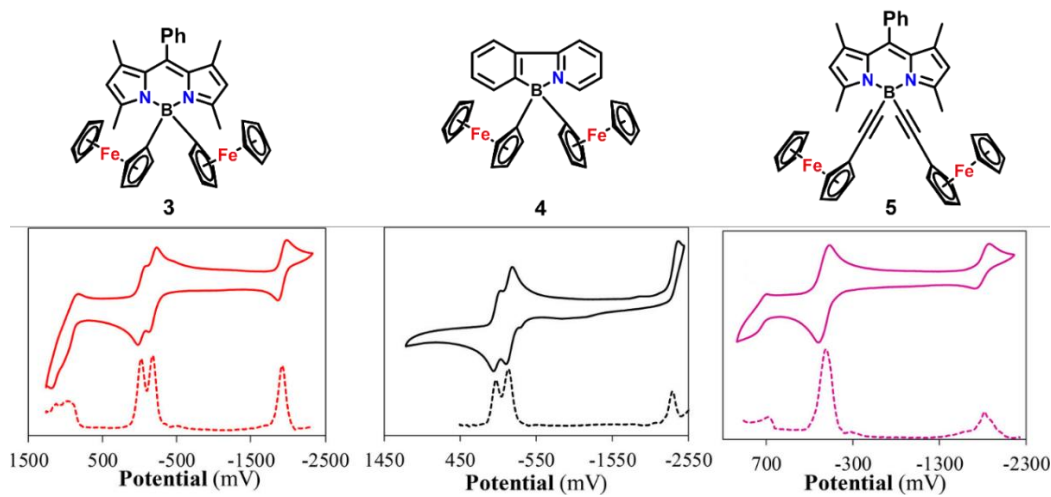

**Fig. 7.** Cyclic and differential-pulse voltammetry of **3**, **4**, and **5**. Reproduced with permission from Ref. [38]. Copyright 2018, American Chemical Society.

## 4. Ferrocene as redox probe for spin-dependent electrochemical charge-transfer process

The effect of the magnetic field in electrochemistry is a well-celebrated topic.[39–44] Seminal theoretical and experimental work observed that magnetic field has altered the charge transfer process.[45–47] The magnetic field-dependence electrochemical reaction involving radical and radical pairs intermediates has also been recently studied.[48,49] To study spin polarization (spin-selective charge propagation) in electrochemistry, ferromagnetic (FM) substrates such as Ni, Co, permalloy are used as working electrodes, which can be magnetized by placing a permanent magnet beneath them.[50–52] Due to the chirality-induced spin selectivity (CISS) effect, chiral molecules coated on magnetic substrates have also been employed to produce spin-selective charge carriers investigated by spin-dependent electrochemical measurements.[53–55] In this respect, achiral ferrocene has been widely utilized as a redox probe to study the magnetic-field effect on electrochemical charge-transfer phenomena for deducing spin polarization (% of SP). Naaman and co-workers grafted a chiral polymer poly{[methyl N-(tert-butoxycarbonyl)-S-3-thienyl-L-cysteinate]-cothiophene} (PCT-L) on ferromagnetic Ni substrates by electrochemical reduction.[56] **Fig. 8a** shows the schematic setup for spin-dependent electrochemical measurements, where PCT-L deposited Ni working electrode was magnetized by the underneath magnet of 0.5 T in UP or DOWN magnetization direction. **Fig 8b** shows the cyclic voltammogram of achiral ferrocene in methanol with PCT-L deposited Ni working electrode as a function of a magnetic field directions. In the case of UP magnetization, current changes its sign from oxidation to reduction. However, in the case of the opposite direction of magnetization (DOWN), it does not vary its direction suggesting a high barrier for charge transfer from PCT-L deposited Ni electrode to $Fc^+$ ion. Overall, researchers successfully demonstrated chiral PCT-L polymer as effective spin filter module allowing only charge carriers with one particular spin as expected in the CISS effect.

To demonstrate on-surface chiral recognition using the chiral polymer, researchers have further recorded cyclic voltammograms on PCT-D and PCT-L coated Au using either the R or S chiral N,N -dimethyl- 1-ferrocenyl-ethylamine as a redox probe. Due to the different chirality of the polymer, there was an enantioselective interaction between deposited polymer and chiral ferrocene redox probe. PCT-D interacted strongly with S chiral N,N -dimethyl- 1-ferrocenyl-ethylamine, whereas PCT-L showed strong interaction with R chiral N,N -dimethyl- 1-ferrocenyl-ethylamine (**Fig. 9**).

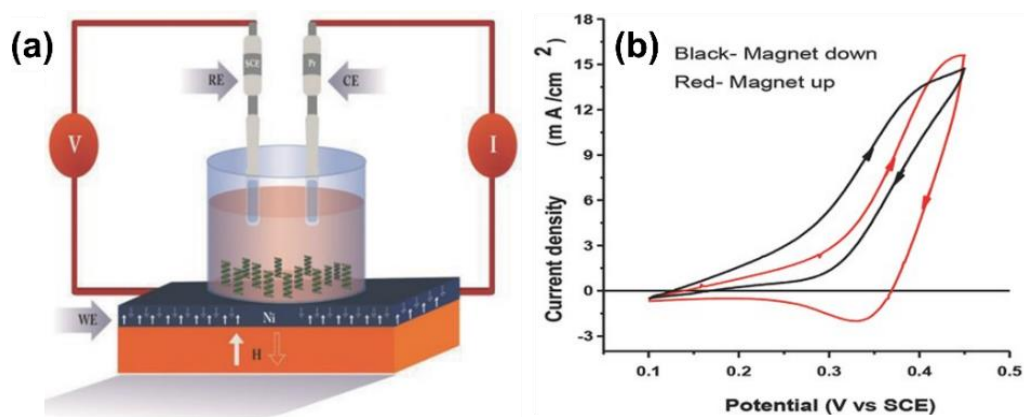

**Fig. 8.** (a) Schematic image of spin-dependent electrochemistry setup, where a permanent magnet was placed beneath PCT-L coated Ni working electrode. (b) Cyclic voltammogram recorded with PCT-L coated Ni electrode in achiral ferrocene as a function of magnetic field. Figures are reproduced with permission from Ref. [56]. Copyright 2015, WILEY-VCH.

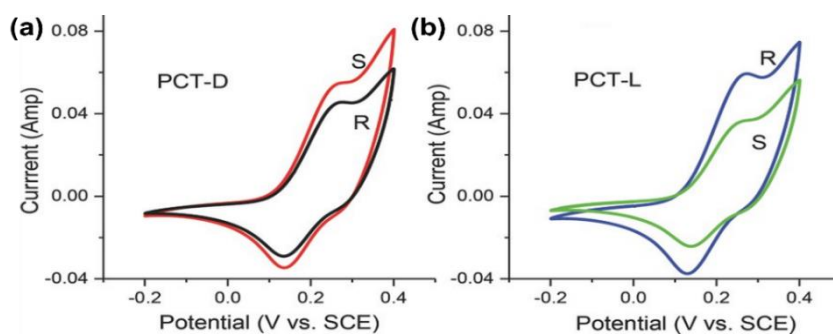

**Fig. 9.** Cyclic Voltammogram recorded with (a) PCT-D Au electrodes in R (black) or S (red) chiral N,N -dimethyl- 1-ferrocenyl-ethylamine. (b) CV recorded with PCT-L Au electrodes in R (blue) or S (green) chiral N,N -dimethyl- 1-ferrocenyl-ethylamine. Figure are reproduced with permission from Ref.[56] Copyright 2015, WILEY-VCH.

Apart from spin-dependent electrochemistry, ferrocene has been employed as redox probe in studying electronic effect of various molecular thin films as well. In this aspect, Charlton *et al.* deposited nitro phenyl as modifier layers on ITO substrates by employing electrochemical reduction of di(4-nitrophenyl) iodonium tetrafluoroborate (DNP) via varying either concentration or number of scans (**Fig. 10**).[57] To investigate, electronic effect of modifier film, CV and EIS was performed using 0.5 mM ferrocene as a redox probe. In the case of NP modified ITO, peak current intensity and peak separation of ferrocene diminished compared to unmodified one due to passivating nature of the NP layer, which depended on DNP concentration and no. of the CV scans. Higher concentration (0.6 mM) and 10 CV scans resulted in grafting of more NP molecules, hence provided more passivation and diminished peak current intensity. EIS study also showed similar electron transfer behavior. **Fig. 10c** shows the Nyquist plot of modified and unmodified ITO in 0.5 mM and 5 mM ferrocene as an electroactive probe. As the concentration increased from 0.2 mM to 0.6 mM, a greater number of NP molecules attached to ITO surface as a result charge transfer resistance ($R_{ct}$) also increased, which was evaluated from bode plot (**Fig. 10d**). Li et al. functionalized glassy carbon (GC) substrate via electrochemical reduction of insitu generated diazonium from 4-(2,5-di-thiophen-2-yl-pyrrol-1-yl)- phenylamine (SNS-An).[58] Ferrocene as redox probes was used to study the electrochemical behavior of SNS-An modified GC substrate (**Fig. 11**). Interestingly, SNS-An modified GC substrate showed a diode like behavior. In modified GC, an irreversible redox behavior of Fc observed. Compared to bare GC, in modified one, FC oxidation peak shifted to 0.2 V vs SCE due to deposited organic layer. It was found that layer behaved as insulating layer below low potential and at higher potential it functioned as conducting like a diode. Using similar diazonium chemistry, Gupta *et al.* modified ITO substrate with naphthyl molecular layers and studied electrochemical behavior of NAPH molecules using similar Fc as redox probe.[27] Compared to bare ITO, the NAPH modified ITO, Fc oxidation peak current density decreased with increase in surface coverage of NAPH molecules illustrating blocking behavior of NAPH films.

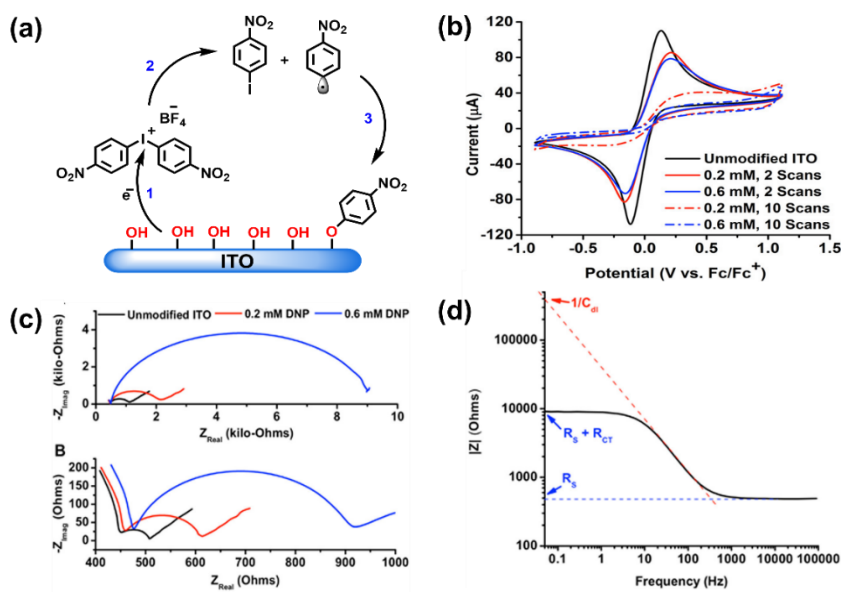

**Fig. 10**. (a) Schematic description of one electron reduction of di(4-nitrophenyl) iodonium tetrafluoroborate (DNP). (b) CVs of unmodified and NP modified ITO in 0.5 mM ferrocene solution. (c) Nyquist plot of unmodified and NP modified ITO in 0.5 mM (above) and 5 mM (bottom) ferrocene solution and (d) its corresponding bode plot in 0.5 mM ferrocene. Figure b, c and d are reproduced with permission from Ref.[57]. Copyright 2015, American Chemical Society.

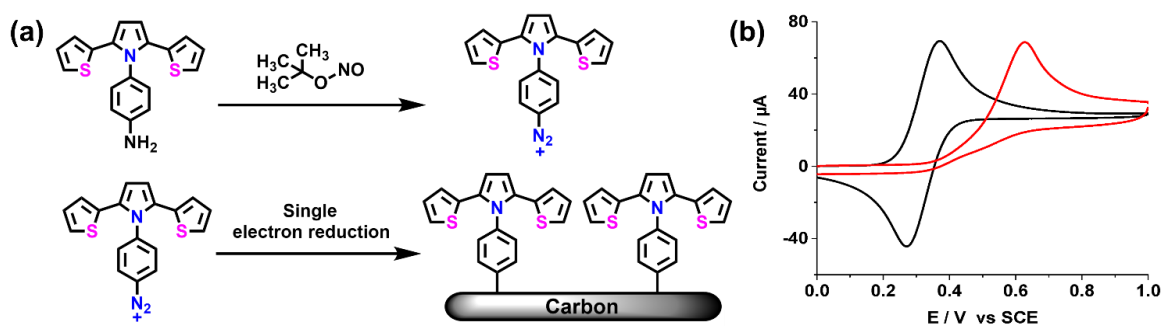

**Fig. 11.** (a) Schematic view of electrochemical grafting of diazonium of 4-(2,5-di-thiophen-2-yl-pyrrol-1-yl)-phenylamine (SNS-An). (b) CVs of bare and modified GC in 1 mM ferrocene redox probe. Reproduced with permission from Ref. [58] Copyright 2020, MDPI.

## 5. On-surface electrochemistry of ferrocene derivatives

Ferrocene has been the subject of focal point to investigate on-surface electrochemistry which is mainly governed by diffusionless charge-transfer process. Many strategies have been utilized to make on-surface assemblies where Fc or its derivatives are attached. For instance, Darwish and co-workers reported covalently bonded monolayers using S-Si interactions and successfully fabricated Fc-containing thin film in much milder conditions.[59] H-terminated Si was employed as a substrate and three different molecules (**8**-**10**) were used to form SAMs. CV studies of SAMs of **8** revealed approximately 0.32 molecule bound with every Si-atom at the surface of Si(111) (**Fig. 12b**), and the scan rate vs. current ($J$) plot show linear relationship indicating diffusion-less charge transfer process (**Fig. 12c**). They used both n- and p-type Si-surfaces for the functionalization and CV measurements which suggested unaffected surface coverage in both the cases (**Fig. 12b & d**). The CV results of **8** SAMs recorded on p-type Si(111)-H surface are comparable with the surface coverage of SAMs on a Au(111) surface (**Fig. 12e**). However, CV data of SAMs **8** prepared on Si(100) (**Fig. 12f**) showed only 33% surface coverage compared to SAMs **8** on Si(111). This unusual value was attributed to incomplete surface H atom replacement. Later, the same group exploited two molecules that contained 1,2-dithiolane-3-pentanoic acid to form SAMs and compared the effectiveness of the process with a third molecule, which was first converted into a disulfide (**Fig. 13**).[60] Spontaneous grafting was performed by dipping Si-H electrode in the acetonitrile solution of **11** or **12**.

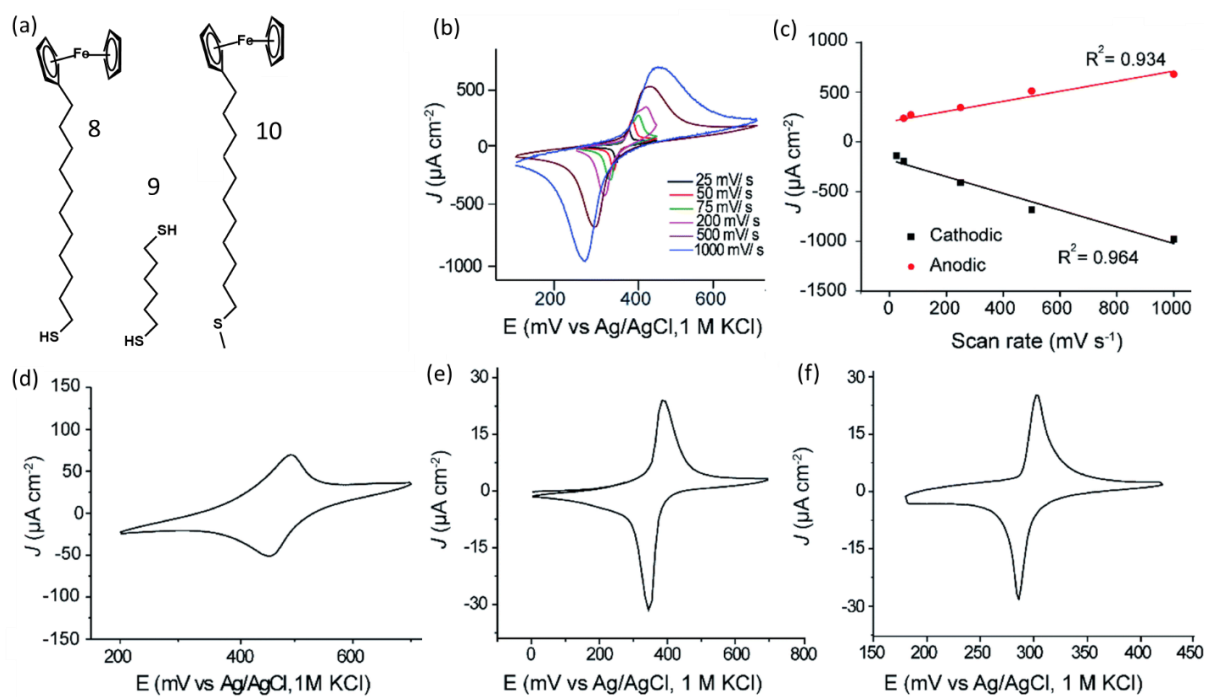

**Fig. 12.** (a) Chemical structure of the compounds 8-10 used for SAMs formation. (b) CVs of SAMs **8** on p-type Si(111)-H surface obtained at different scanning rates, and (c) linear plot of current vs scan rates. (d) CV of SAM **8** on a n-type Si(111)-H surface, (e) CV of SAM **8** on Au(111) surface, (f) CV of SAM **8** on H-terminated Si(100). Reproduced with permission from Ref. [59] Copyright 2020, Royal Society of Chemistry.

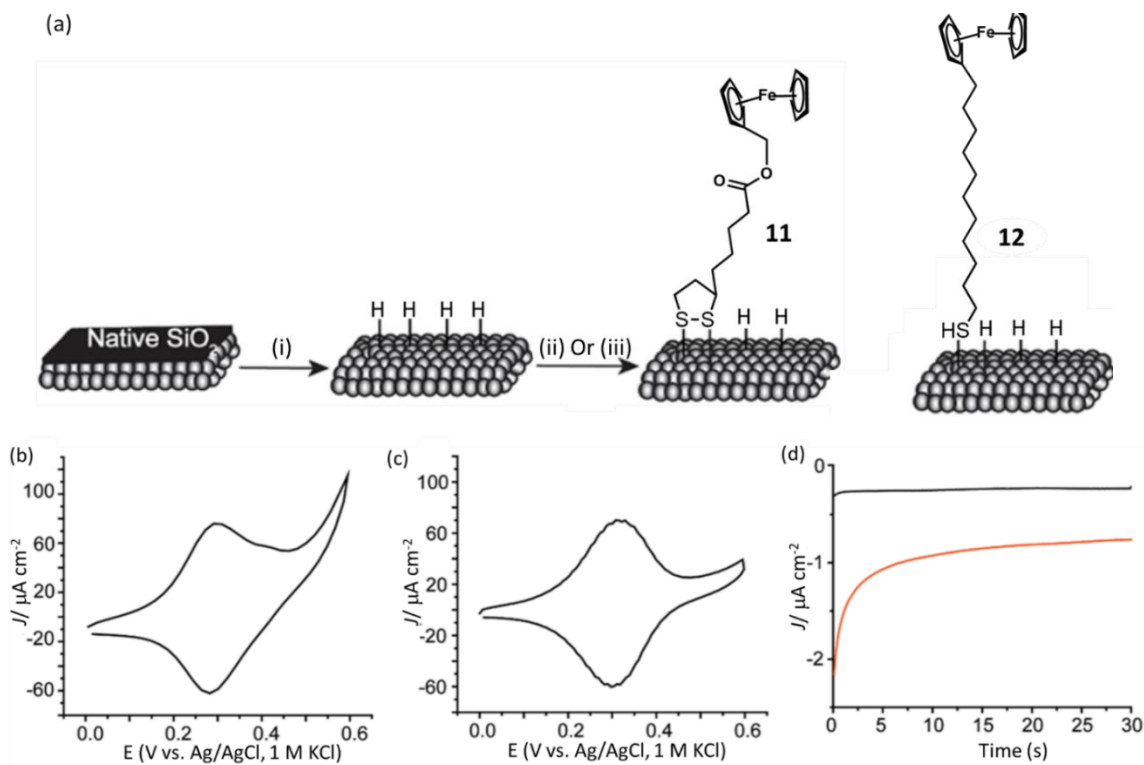

**Fig. 13.** (a) Scheme for preparation of SAMs **11** and **12** on Si surface. CVs of SAMs **11** on Si surface obtained by spontaneous grafting method (b), and electrochemical grafting (c). (d) current density vs. time plot obtained from electrochemically grafted SAMs **11** using potential 0.8 V (black line) and at 0.9 V (red line). Reproduced with permission from Ref. [60] Copyright 2020, American Chemical Society.

To have a clearer view about presence of $O_2$, an argon bubbled solution of **11** was used to see if SAMs formation takes place. Cyclic voltammograms support the fact that presence of oxygen is necessary. Conversion of thiol into disulfide basically reduced the monolayer formation time in ambient conditions from 24 h to 1 h. A comparative study performed by preparing monolayer of **11** on Si and Au both revealed that to reach comparable surface monolayer coverage, Au needed 2 h more than the Si-surface. This study could be extremely helpful in preparation of protein functionalized Si-surfaces. SAMs on semiconducting surface such as on Si also functions as a protecting layer which stops the reoxidation process of freshly prepared Si-H surface. For this reason, after removal of the insulating oxide layer from Si-surface the SAMs formation needs to be established rapidly while maintaining the oxygen free condition. However, studies have shown that even in highly dense alkyne-based SAMs the coverage area is up to 60-65% and therefore nearly 35% area remains prone to reoxidation allowing growth of insulating oxide regions on the surface.[61,62] The same group investigated the effects of such unavoidable oxide growths on the Si-based SAMs and the charge transfer kinetics using electrochemical impedance spectroscopy (EIS) (**Fig. 14**)[63] In their study, they followed two procedures. In one procedure ferrocene terminated SAM was maintained under ambient condition and charge transfer kinetics were correlated with the exposure time. In the second approach, the terminal alkyne functionalized SAM was first left in ambient conditions for a certain amount of time and then ferrocene group was introduced via a click reaction and then charge transfer kinetics was studied (**Fig. 14b**). Oxidation and reduction peak current comparisons for day 1 and for day 30 showed a decrease indicating a corresponding decrease in surface coverage area. The estimated surface coverage area was found to be nearly 37% less than day 1. Peak separation also increased from 50 mV to 140 mV in 30 days suggesting slower kinetics for electron transfer process. EIS measurements of S-2 revealed an increase in resistance with the increasing exposure time (**Fig. 14c**). Both measurements exhibit a decrease in surface coverage area with increasing exposure time. Fact that surface coverage area decrease is caused by oxidation and not due to ferrocene decomposition, was investigate by performing control experiment.

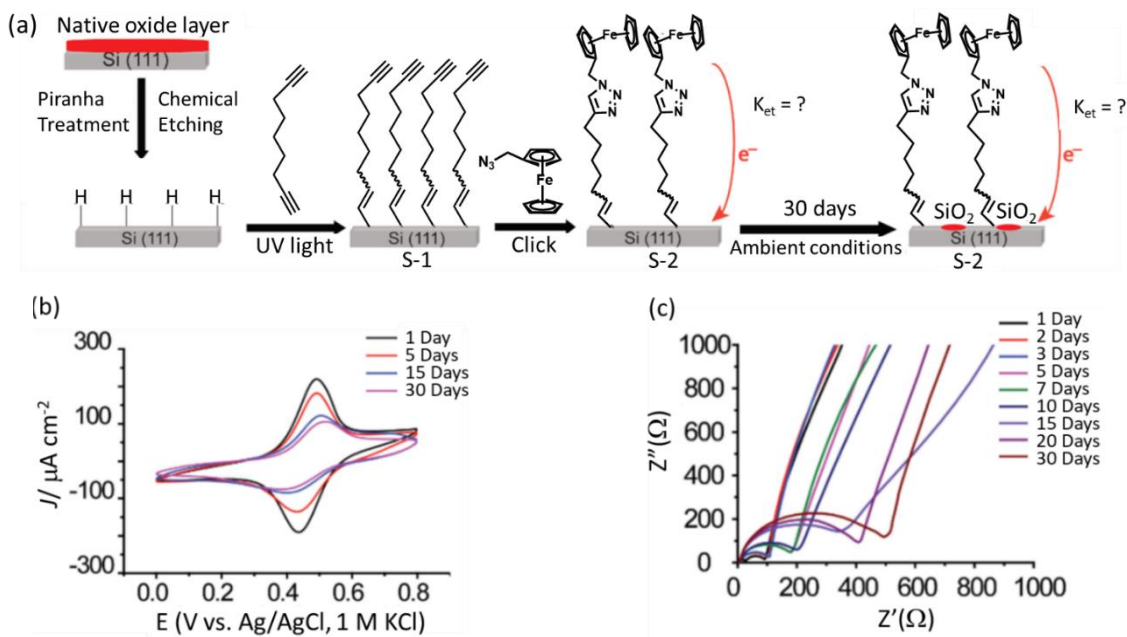

**Fig. 14.** (a) Schematic presentation of the SAMs H-terminated Si(111) substrates. (b) CV measurements on a SAMs S-2 sample at day 1, 5, 15, and 30. (c) EIS measurement corresponding to sample exposed for different amount of time (1 day-30 days). Reproduced with permission from Ref. [63] Copyright 2021, American Chemical Society.

Fontanesi and co-workers prepared ferrocene monolayers using -OH group of a Fc-derivative that was covalently grafted on H-terminated Si(111) surface (**Fig. 15**).[64] CV measurements performed using Si-surface functionalized with methanol-ferrocene (Me-Fc) and 1-iodoundecanoic acid-ferrocene (UA-Fc) revealed a quasi-reversible oxidation for grafted ferrocene. The calculated surface coverage for Si-Me-Fc monolayer ($4.3 \times 10^{-10}$ mol/cm$^2$) indicated a nicely packed monolayer structure. Si-UA-Fc yielded similar CV characteristics; however, the reversibility was even less when compared to Si-Me-Fc.

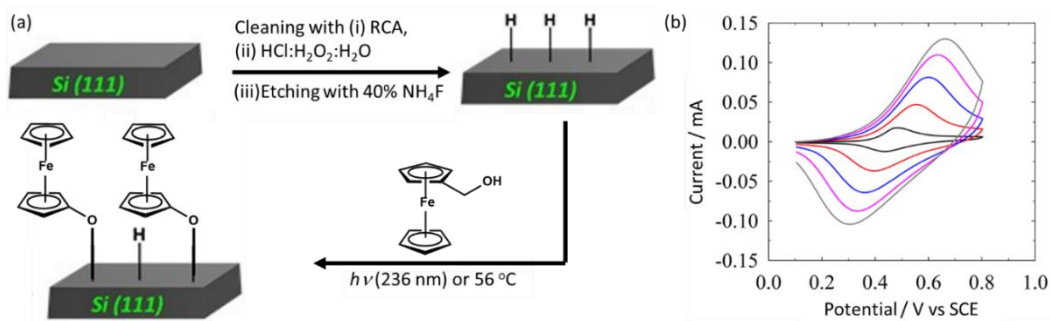

**Fig. 15.** (a) Scheme for surface modification process of the H-terminated silicon substrates. (b) corresponding CVs. Reproduced with permission from Ref. [64] Copyright 2019, Springer Nature.

Taherinia prepared a range of ferrocene terminated SAMs on a gold surface by varying oligophenyleneimine spacers and investigated the interfacial electron transfer kinetics (**Fig. 16**).[65] The SAMs were obtained in a step-wise imination method. First the Au-substrate was soaked in ethanol solution of 4-aminothiophenol to get amin-terminated SAMs which was then reacted with a terphthalaldehyde solution in ethanol resulting in an imine product. In the next step this was further reacted with a diamine followed by another reaction with dialdehyde. The SAMs were employed in electrochemical measurements recorded in ionic liquid as an electrolyte and show an increase in peak separation upon addition of cyclohexyl group. This separation was not observed when 0.1 M $Bu_4NPF_6$ in acetonitrile was used for the CV measurements which was attributed to the difference in SAMs and electrolyte interactions. With increasing temperature, the decrease in peak separation was observed in all SAMs. Plot of electron transfer rate (ln k°) against molecular length for SAMs of OPI 2_Fc, OPI 4_Fc and OPI 8_Fc show a linear trend indicating tunneling charge transport.

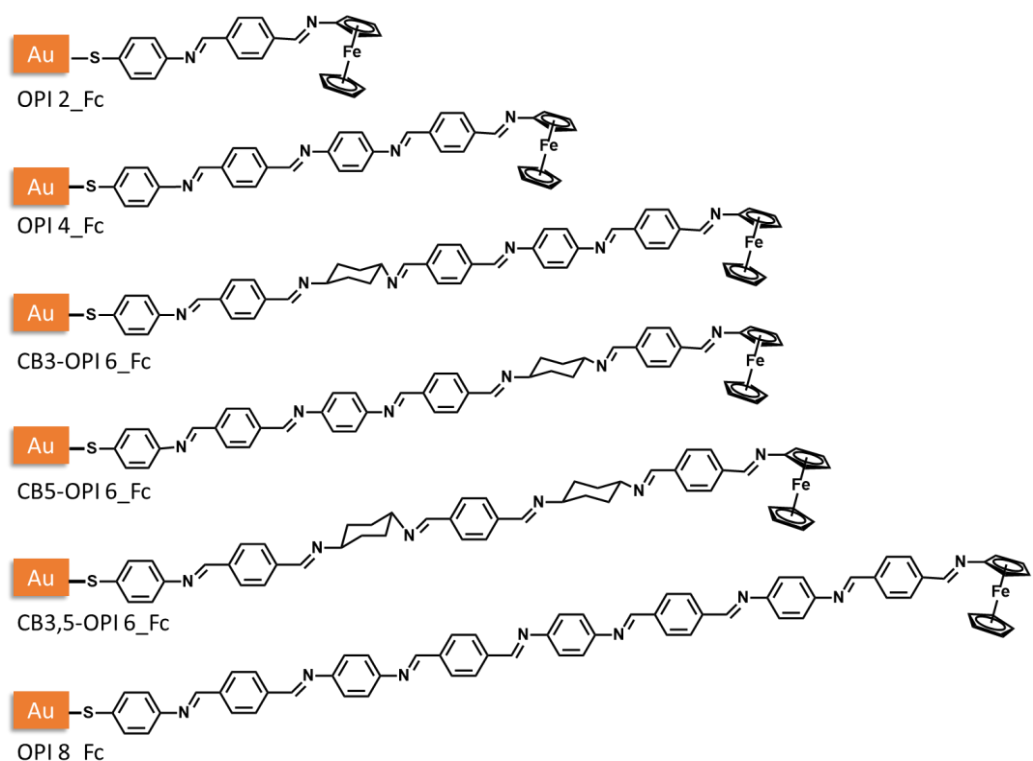

**Fig. 16.** Scheme showing ferrocene-terminated SAMs. Reproduced with permission from Ref. [65] Copyright 2020, American Chemical Society.

Functionalization of carbon electrodes is another area where ferrocene-based SAMs have been used.[66] In a typical procedure the glassy carbon electrode is modified with amine functional SAMs and then reacted with ferrocene-carboxylic acid in presence of coupling reagents EDC and NHS (**Fig. 17a**).[67] CV measurements showed that when dilution was 50 mol % the peaks became sharper compared to SAMs prepared with no diluent (**Fig. 17b**). This is probably due to possible intermolecular electron transfer between two close ferrocene units resulting in broadened peaks which could be avoided in diluted sample. Dilution beyond 50 mol% results in decreased peak current densities due to the reduced surface coverage. The ferrocene interaction in undiluted sample was also obvious from the CV peak separation and the full width at half maximum values which were higher than those obtained on Au-surface. Ferrocene-SAMs having six and twelve $CH_2$ group spacers and studied its effect on surface coverage. CV measurements exhibited similar peak separation, however, longer-chain SAMs followed slower electron transfer kinetics. It is known that carbon electrode surface is less defined than the Au and contains more defect sites. It was proposed that alkyl chains in SAMs may bend bringing ferrocene head to be in close contact with the defect sites. Also, a larger difference in spacer lengths between ferrocene linker and diluent would again cause ferrocene-ferrocene interactions leading to broadening of peaks. To clear these doubts, they prepared SAMs having same linker (C6 and C12) and diluent spacer ($NH_2(CH_2)_nOH$, n=6 & 12). Comparative studies revealed that similar length diluents cause significant reduction in ferrocene attachments probably due to steric hindrance in amidation, and also results in much less non-Faradic current densities. Additionally, they prepared SAMs in which defect sites were blocked by electrodeposition of ZnO (**Fig. 18**).

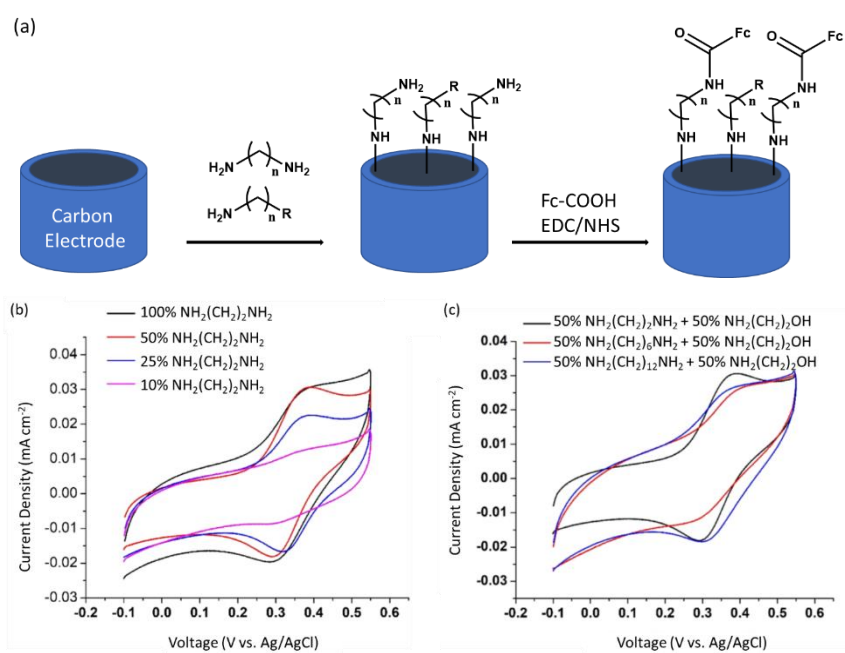

**Fig. 17.** (a) Scheme showing carbon electrode modification using ferrocene derivative. (b), (c) CVs of ferrocene containing SAMs prepared in different conditions. Reproduced with permission from Ref. [67] Copyright 2021, American Chemical Society.

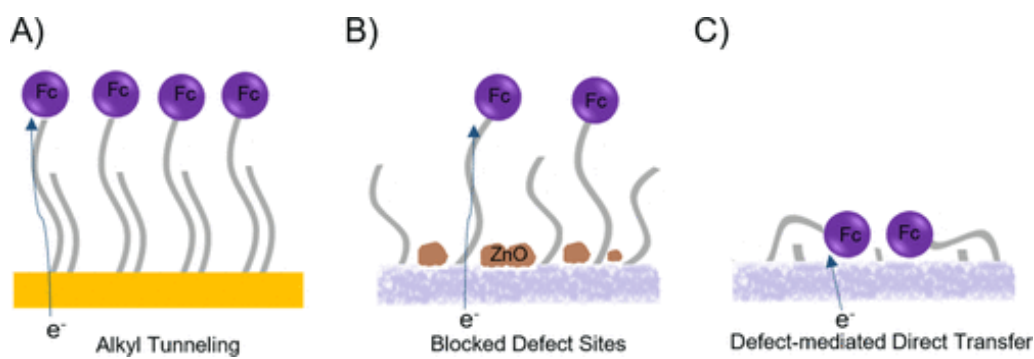

**Fig. 18.** (a) Diagram showing electron tunneling scheme in ferrocene SAMs on gold surface, (b) use of ZnO to block defect sites on carbon electrode, (c) charge transfer in absences of ZnO coating. Reproduced with permission from Ref. [67] Copyright 2021, American Chemical Society.

Another interesting work describing the packing and molecular orientations in a ferrocene SAMs obtained by performing 'click chemistry' was reported by Escoubas and co-workers.[68] They have applied three different synthetic strategies to prepare the ferrocene SAMs. In one strategy they did one step direct deposition, in second and third, they used two-step process employing click reaction between alkyne and azide functionalities (**Fig. 19**). SAMs preparation by direct deposition was performed by soaking a fresh template stripped Au substrate in a 1 mM solution of ferrocene functionalized compounds. For the two-step procedure, first substrate was soaked for 20 h in 1 mM solution of mixed thiol (containing diluent octanethiol in 1:1 ratio) and then click reaction with ferrocene unit having alkyne or azide funcionalities was performed. CVs were recorded in 1 M aqueous $HClO_4$ and using Ag/AgCl reference electrode was used to characterize the various SAMs. The deconvolution analysis of the obtained voltammograms from SAMs **16** yielded peak areas $E_{pa,I}$ and $E_{pa,II}$ as 72% and 28% respectively, which were attributed to the two different arrangements of Fc units corresponding to standing-up phase and buried phase in the SAMs. Surface coverage area calculation showed

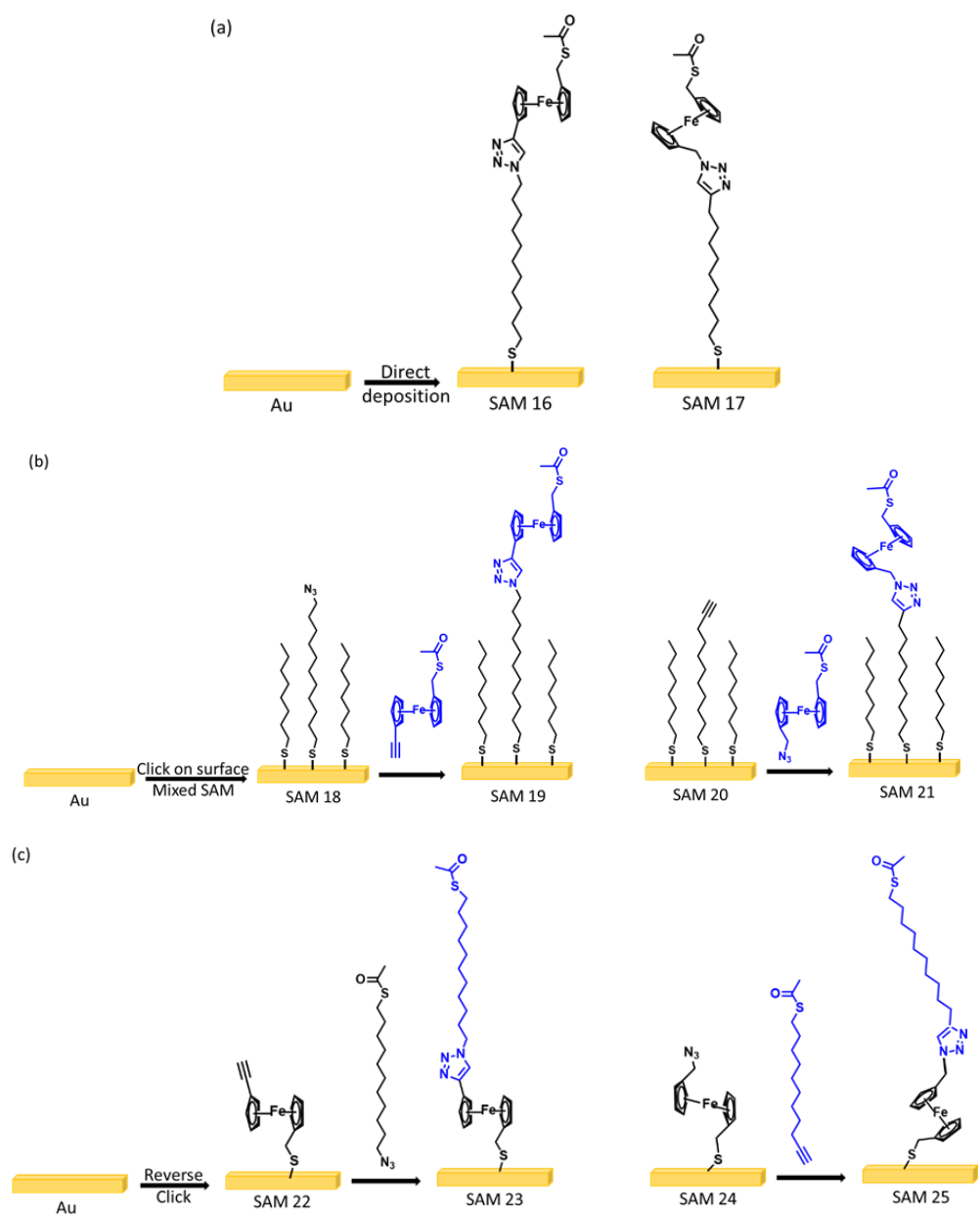

**Fig. 19.** Schematic diagram showing various synthetic strategies used to prepare of SAMs **16**-**25**. Reproduced with permission from Ref. [68] Copyright 2022, American Chemical Society.

57% and 100% coverage values for SAM **16** and SAM **17**, respectively. This difference can not arise due to the spacer nature (odd-even effect) as both molecules have odd number of carbon atoms in the chain. Authors proposed that the strain between S- and Fc units leads to bending of molecule which effectively reduced the available surface in SAMs **16**, whereas, in SAMs **17** the spacer between triazole and Fc unit help in adjusting the strain. Kim and co-workers observed electronic and interfacial structural changes in a Fc-SAMs prepared on Au.[69] To probe such an exciting phenomena, they clubbed XPS, UPS along with a conventional 3-electrodes based electrochemical set-up for measuring thickness of the Fc-SAMs during the redox events (**Fig. 20**). They elucidated that during an oxidation process, when Fc is converted to $Fc^+$, due to strong electrostatic interactions between the oxidized redox probe and electrolyte ($ClO_4^-$), facilitate the SAMs to be nearly perpendicular on the underlying Au substrate. As a result, thickness of the $Fc^+$-SAMs increases to $18.6 \pm 0.2$Å as compared to the neutral Fc-SAMs which was at $18.6 \pm 0.2$Å. Besides, they experimentally deduced a change in the dipole moment due to formation of strong ion-pairs ($Fc^+$- $ClO_4^-$) by measuring the work function.

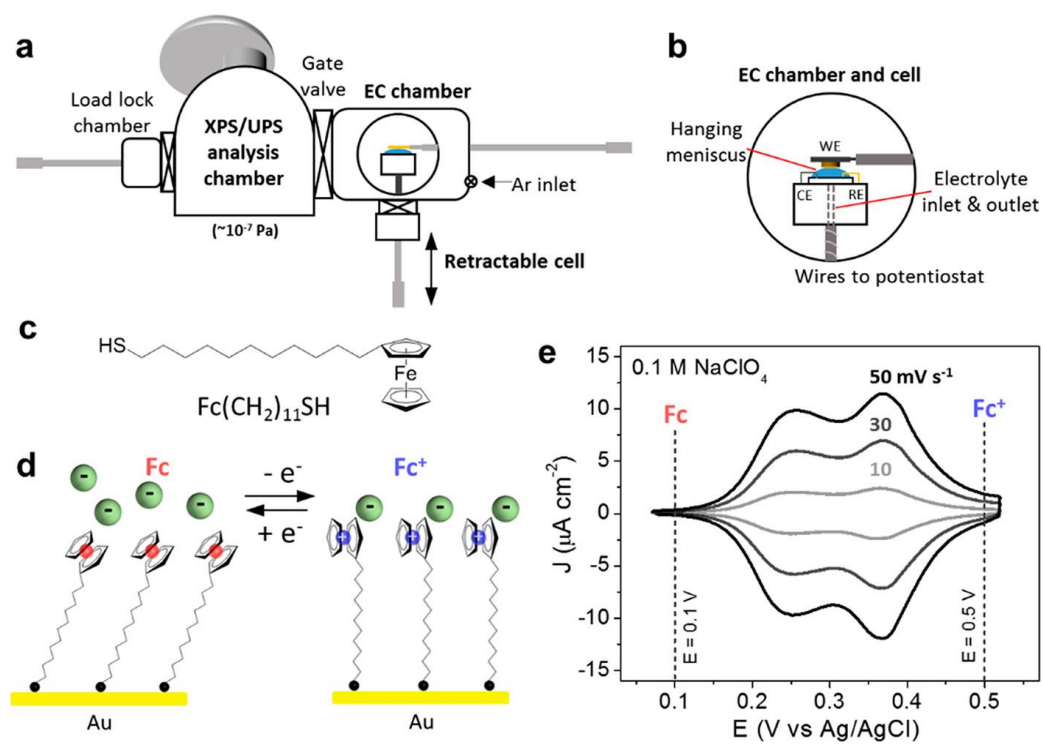

**Fig. 20.** (a-b) X-ray and UPS clubbed with electrochemical measurements set-up, (c) chemical structure of Fc-thiolated SAMs, (d) schematic description of electrode/SAMs interfacial structural changes during electrochemical oxidation and reduction, (e) CVs recorded on Au/Fc SAMs at various scan rates. Reproduced with permission from Ref. [69] Copyright 2018, American Chemical Society.

## 6. Photo-driven charge transfer

Systems in which charge transfer could be triggered by light absorption are of special interest. One of the major areas where such materials find application is the field of solar cells. Ferrocene-based complexes have been employed in dye sensitized solar cells (DSSC) and their ability as sensitizer in TiO$_2$ based solar cells been investigated in past.[70–73] In the context of solar cells, titanium-oxo-clusters, a family of polynuclear titanium compounds, have seen growing interest towards their possible roles in improving the efficiency.[74,75] Number of strategies have been adopted to tailor the optical properties of titanium-oxo-clusters among which the most frequently applied route is the synthesis of heterometallic TOCs which helps in reducing the band gap and shifting the absorption to visible light region.[76,77] Fan and co-workers realised that interesting changes in the optical properties of titanium-oxo-clusters could be obtained upon the introduction of ferrocene moieties.[78] They prepared such clusters that are substituted with ferrocenecarboxylate [Ti$_6$(μ$_3$-O)$_6$(O$^i$Pr)$_6$(OOCFc)$_6$] and compared it with structurally similar derivatives obtained by using 1-naptholate (NA) [Ti$_6$(μ$_3$-O)$_6$(O$i$Pr)$_6$(NA)$_6$] and 1-naphthylacetate (NAA) [Ti$_6$(μ$_3$-O)$_6$(O$i$Pr)$_6$(NAA)$_6$], respectively (**Fig. 21**). UV-vis absorption spectra of naphthyl-functionalised clusters show absorption below 400 nm while ferrocene derivatised clusters exhibits absorption extending up to 600 nm. Calculated band gaps were found to be 2.2, 3.2 and 3.3 eV for Fc, NA and NAA-functionalised cluster, respectively. The strikingly lower band gap of Fc-functionalised TOC was attributed to Fe (II) d-d transitions and presence of some Fc-{Ti$_6$} charge transfer. The CV of the Fc-functionalized cluster performed in dichloromethane using supporting electrolyte n-Bu$_4$NPF$_6$ showed a quasi-reversible redox behaviour with E$_{pa}$=0.723 and E$_{pc}$=0.616 V (**Fig. 22a**). Appearance of just one peak is indicative of the fact that the six Fc units are independent of each other and undergo a simultaneous oxidization process. In their study they found that the CV of the starting material ferrocenecarboxylic acid exhibited a completely irreversible oxidation peak at 0.78V. The contrast change in the redox reversibility of the Fc units in cluster compound indicated that there is some electronic interaction between ferrocenecarboxylate ligand and the {Ti$_6$} moiety. DFT calculations revealed that the two highest occupied molecular orbitals (HOMOs) have significant contribution from the Fe 3d orbitals whereas the two lowest unoccupied molecular orbitals (LUMOs) are located on the Titanium 3d orbitals of {Ti$_6$} cluster. Photoelectrochemical properties of clusters showed anodic current suggesting n-type semiconduction in the clusters under study (**Fig. 22b**). While all the three tested clusters exhibited a steady and reproducible photo current during many on/off cycles of Xenon light radiation, cluster functionalised with Fc units showed significantly better photocurrent implying efficient electron injection from excited states of Fc sensitizers.

Dai group prepared a similar cluster functionalized with multiferroic units, [Ti$_6$O$_6$(O$^i$Pr)$_6$(O$_2$CFc)$_6$] and compared its potential as electrochemical sensor for sugar reduction using generated photocurrent.[79] A structurally similar compounds that did not contain ferrocene [Ti$_6$O$_6$(O$^i$Pr)$_6$(O$_2$C$^i$Bu)$_6$] was used as reference (**Fig. 23a-b**). The voltammograms reveal simultaneous oxidation of all the six Fc units showing a reversible redox potential with a half-wave redox potential at E$_{1/2}$ = 0.62 V which is slightly larger than the value observed for free ferrocene (E$_{1/2}$=0.558 V) (**Fig. 23c**). This difference might be attributed to the withdrawal of electron density by Ti(IV) centre in the molecule. They observed that Faradaic current varied upon addition of glucose and noticed a significant current decrease with increasing time (**Fig. 23d**). Photocurrent response carried out in 0.1 ml L$^{-1}$ sodium sulphate electrolyte solution under illumination from a 150-W xenon lamp revealed about three times higher photocurrent densities than non-Fc cluster. The sensitivity of Fc-functionalised cluster for four saccharides glucose, fructose, maltose and sucrose was tested. A linear dependence in current density and added glucose was observed. While it is important to optimize the molecular designs and fabrication strategies that may improve the efficiency of any system, same time this is also of equal importance to know the things that must be avoided. Koivisto and co-workers studied the effect of 4,4-difluoro-4-bora-3a,4a-diaza-s-indacenes (BODIPY)-ferrocene dyads on DSSC and observed some unfavourable electron transfer in electron deficient BODIPY-ferrocene when in conjugation.[80] They prepared β-substituted Fc-BODIPY derivatives and performed the detailed investigation. UV-vis absorption studies showed much broader absorption band for Fc-functionalized BODIPY compound compared to system lacking a Fc unit. This feature is desirable in photovoltaic as it means absorption of a wide range of light spectrum and hence better efficiency of the device. However, cyclic voltametric investigations were discouraging. BODIPY cores in its inherent state exhibit pseudo reversible reduction but when it is functionalised with ferrocene the reversibility gets lost, irrespective of the meso-position substituent's nature. Ferrocene-BODIPY dyads did exhibit one electron reversible oxidation. Observed instability in Fc-BODIPY is different from previous observations on similar system as in those cases Fc moiety was the cause, however, in present system it is BODIPY site that is decomposing. This study eliminates the existing doubts about Fc-BODIPY dyads and indicates that the either conjugation between two redox sites should be broken or α-position of BODIPY should be modified in order to elevate the energy of the LUMO and thereby stop the ground state electron transfer.

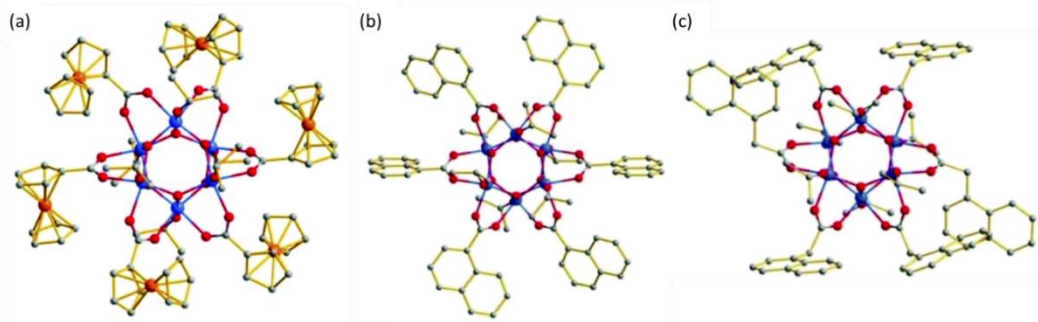

**Fig. 21.** Single crystal structure of titanium-oxo-clusters containing ferrocene moiety [Ti$_6$($\mu_3$-O)$_6$(OiPr)$_6$(OOCFc)$_6$], and structurally similar [Ti$_6$($\mu_3$-O)$_6$(OiPr)$_6$(NA)$_6$] and [Ti$_6$($\mu_3$-O)$_6$(OiPr)$_6$(NAA)$_6$] lacking the redox unit. Reproduced with permission from Ref. [78] Copyright 2017, Royal Society of Chemistry.

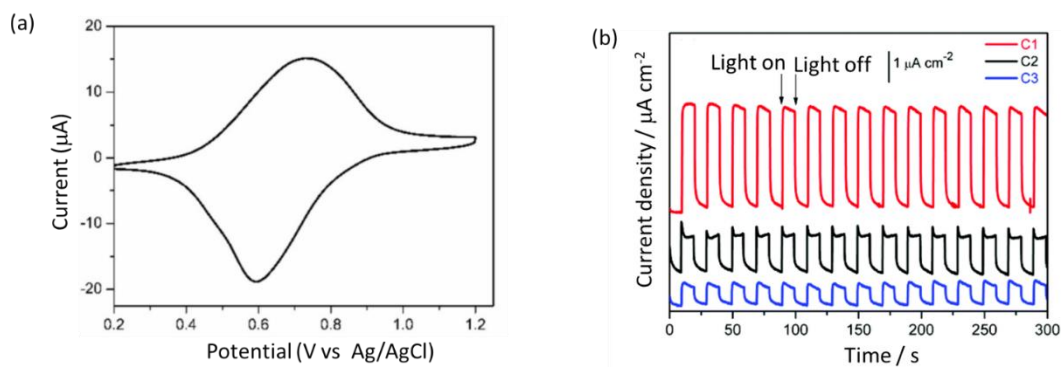

**Fig. 22.** (a) CV of Fc-functionalized cluster recorded at 100 mV/s. (b) photocurrent measurements for NA (black), NAA (blue) and Fc-functionalized clusters (red). Reproduced with permission from Ref. [78] Copyright 2017, Royal Society of Chemistry.

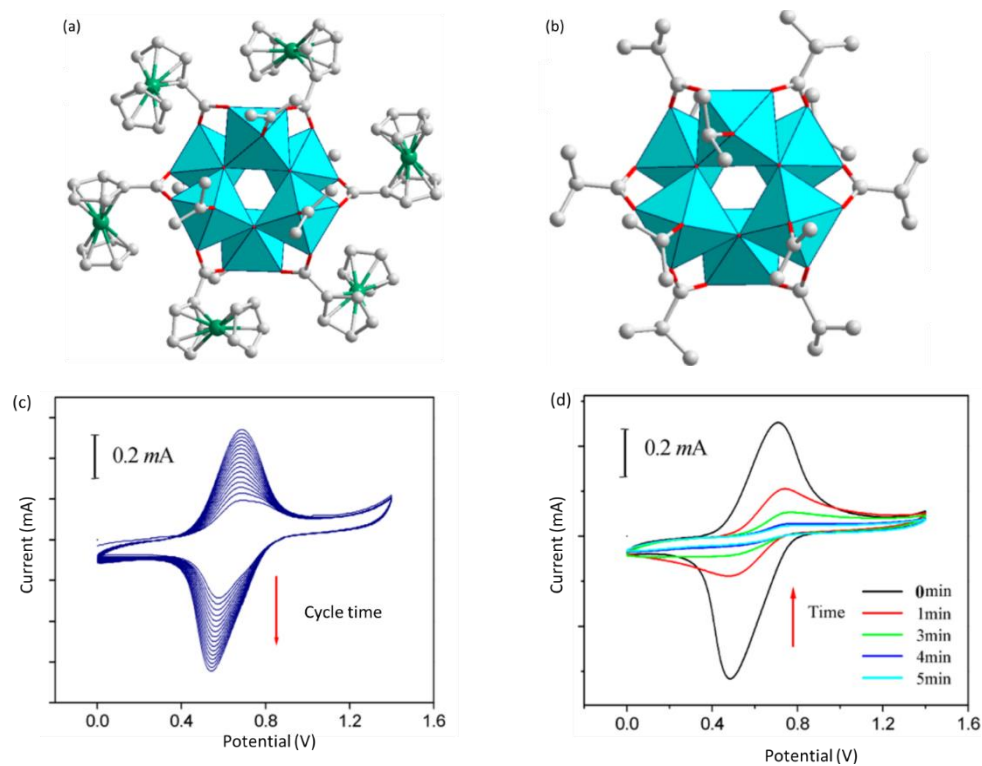

**Fig. 23.** (a) Solid state structure of ferrocene functionalized cluster, $[Ti_6O_6(O^iPr)_6(O_2CFc)_6]$, (b) without ferrocene. (c & d) CVs of $[Ti_6O_6(O^iPr)_6(O_2CFc)_6]$ in dichloromethane and upon addition of glucose (2 mg) at intervals of 1 minute. Reproduced with permission from Ref. [79] Copyright 2017, American Chemical Society.

Sun and co-workers employed ferrocene as volatile solid additive and explored a new way to optimize the bulk heterojunction (BHJ) morphology in organic solar cells (OSC).[81] In general, during OSC device fabrications high boiling point solvent additives are used which removal becomes difficult and may cause unwanted variations in morphology affecting performance and reproducibility negatively. They observed one of the highest power conversion efficiencies (17.4%) and short circuit current density ($J_{sc}$) of 26.9 mA cm$^{-2}$ with a ferrocene treated system. Interestingly, device without any additive showed PCE of 15.5% and one treated with CN$^-$ solution yielded 16.5%. The improvement in observed properties is probably due to better molecular crystallinity, higher transport and suppressed recombination in ferrocene processed device. Bera and co-worker prepared ferrocene-based donor-acceptor type ferrocenylcyanovinyl compound in a solvent-free method with the idea to create a system that serves as dye and redox electrolyte (**Fig. 24**).[82] They fabricated an electrolyte-free DSSC and obtained an open circuit voltage of 0.84 V much higher than the value obtained with conventional DSSC which was used to measure photovoltaic properties using 100 mW/cm$^2$ illumination is shown in **Fig. 24b**. The group observed short circuit photocurrent density ($J_{sc}$) of 0.41 mA/cm$^2$ and 39 μA/cm$^2$ for device having different chemical compositions.

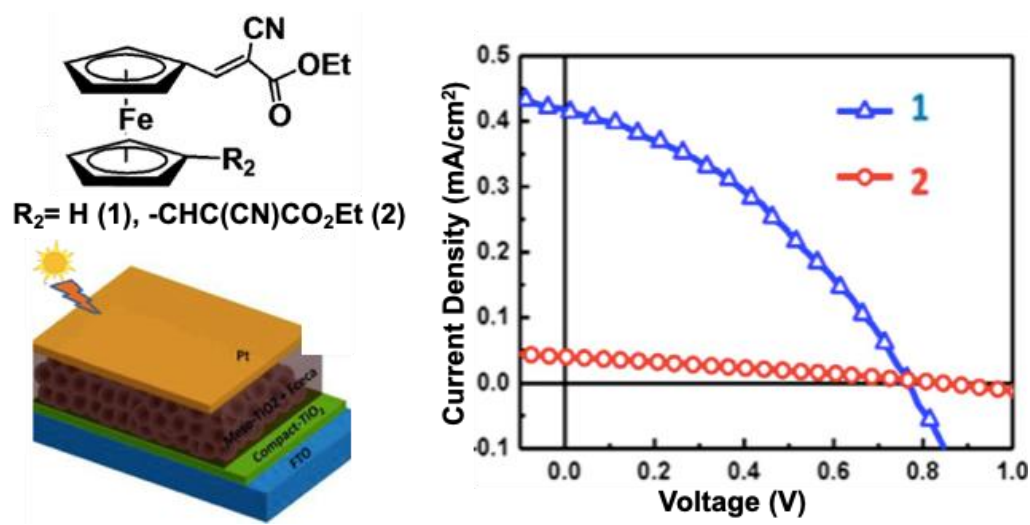

**Fig. 24.** (a) Chemical structure of studied ferrocene system, schematic diagram of prepared device, (b) J-V characteristics of the heterojunction solar cells under light. Reproduced with permission from Ref. [82] Copyright 2018, American Chemical Society.

As beautifully described by the Lalevée and co-workers, photoredox catalysis in presence of near infrared (NIR) remains highly desirable but quite difficult to achieve due to relatively much lower energies of NIR photons.[83] They explored the possibility of using ferrocene derivative towards photopolymerization reaction. The idea was to use the photooxidation of Fe(II) in ferrocene and let excited iron to react with an iodonium salt ($Ar_2I^+$). Polymerization reactions revealed that under mild irradiation (405 nm) all the ferrocene/iodinium compounds Fe/Iod, Fe5/Iod, Fe3/Iod, Fe4/Iod show slow polymerisation and no activity could be observed for Fe6 & Fe7. However, when 2-diphenylphosphinobenzoic acid (2-dppba) was introduced, a huge jump in radical polymerization rate was observed. Fe(II)*/$Ar_2I^+$ couple was then used to initiate photopolymerization under 78 nm light. In NIR light, Fe4/Iod showed some activity whereas Fe7Me/Iod did not show any. Increasing the Fe(II) from 0.1wt% to 0.5 wt% in Fe4 system improved the efficiency significantly. Another big challenge that world is facing now is the extreme emission of greenhouse gases, predominantly comprising of carbon dioxide. A huge amount of research has been dedicated to address this alarming environmental issue. In this context, large number of photocatalysts that may reduce the $CO_2$ into a reusable fuel has been investigated in past. Lan group explored the ferrocene-functionalised polyoxo-titatnium clusters (PTCs) for photoreduction of $CO_2$.[84] In order to modulate the absorption, charge transfer properties and to improve photocatalytic activity towards $CO_2$ reduction, three derivatives **Ti$_6$**, **Ti$_8$-Fcdc** and **Ti$_6$-Fcdc** having molecular formula as [Ti$_6$($\mu_3$-O)$_2$($\mu_2$-O)$_2$(PPO)$_2$(O$_2$CCF$_3$)$_2$($\mu_2$-O$^i$Pr)$_4$(O$^i$Pr)$_6$], [Ti$_8$($\mu_3$-O)$_4$(tea)$_2$((Fcdc)$_2$)$_4$(O$^i$Pr)$_{10}$]·2HO$^i$Pr and [Ti$_6$($\mu_3$-O)$_2$((Fcdc)$_2$($\mu_2$-SO$_4$)$_2$($\mu_2$-O$^i$Pr)$_2$(O$^i$Pr)$_{10}$]·2HO$^i$Pr where, CF$_3$COOH = trifluoroacetic acid, HO$^i$Pr = isopropanol, teaH$_3$ = Triethanolamine, and Fcdc = 1,1-ferrocene dicarboxylic acid (**Fig. 25 a-c**) were prepared. The incorporation of Fcdc group in the molecule extended the absorption band up to 600 nm in the visible region of light spectrum. The photocatalytic $CO_2$ reduction activities of **Ti₆, Ti₈-Fcdc** and **Ti₆-Fcdc** were evaluated under visible light illumination showing formation of $HCOO^-$ as a function of time (**Fig. 25d,e**). A continuous increase in $HCOO^-$ concentration could be seen in 10 h of reaction time reaching up to 4.03, 17.03 and 35.00 μmol for **Ti₆, Ti₈-Fcdc** and **Ti₆-Fcdc,** respectively. Higher activity and selectivity at photocatalytic reduction of $CO_2$ of **Ti₈-Fcdc** and **Ti₆-Fcdc** over **Ti₆**, is attributed to efficient charge transfer by Fcdc ligand. Among the two Fcdc derivatives studied here, **Ti₆-Fcdc** exhibited better photocatalytic performance than **Ti₈-Fcdc**. Therefore, having ferrocene in the PTC system could effectively catalyse the $CO_2$ reduction in water that too under visible light. Also, the catalytic performance exhibited by **Ti₆-Fcdc** is highest among known PTC-based catalysts.

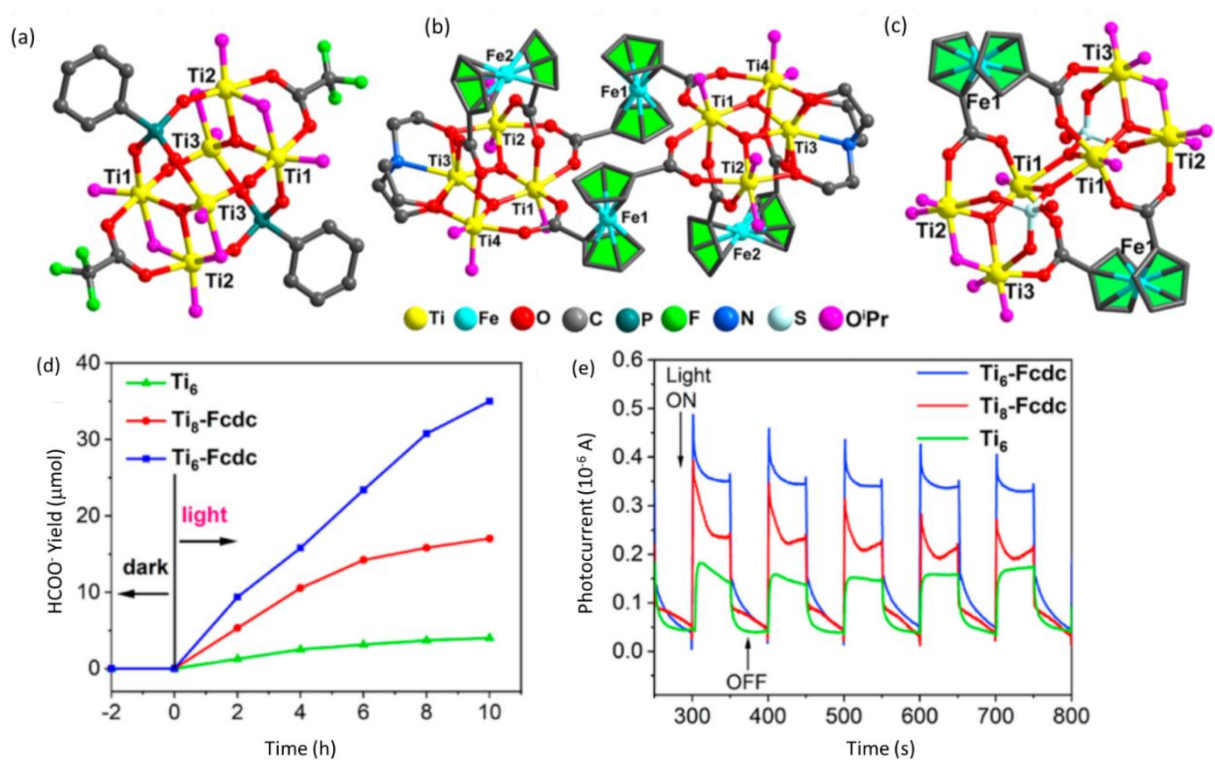

**Fig. 25.** (a-c) Solid state molecular structure of **Ti₆, Ti₈-Fcdc** and **Ti₆-Fcdc**. (d & e) show the yield of $HCOO^-$ and (b) showing the photocurrent measurement of Ti₆, Ti₈−Fcdc, and Ti₆−Fcdc under visible light irradiation. Reproduced with permission from Ref. [84] Copyright 2021, American Chemical Society.

## 7. Ferrocenes in molecular electronics

Over the last few years, the scientific community have developed an intricate understanding of the correlation between chemical and electronic parameters of molecular electronic devices.[85–88] These findings have boosted the research activity around molecular electronics which utilises a single molecule as an active part of the electronics circuit.[100–103] The main aim of molecular electronics is to reduce the size of the device to the molecular scale (few nm) thereby, increasing the packing density of the device. Single molecule junctions are comprised of three characteristic parts namely-anchor, electrode and a bridge.[92,93] Fast and reversible redox reactions involving the $Fc/Fc^+$ couple makes it a suitable candidate in electronic devices.[94,95] Once active molecular unit is prepared, the next challenge that is faced is the fabrication of mechanically robust contacts that connects the molecule mostly in sandwiched configuration.[96] The most widely used electrode-molecule contacts have been based on the chemisorption of thiols on gold electrode.[97–99] However, due to certain drawbacks such as lability of the S-Au bonds and changes in structure due to induced electric field, semiconducting platforms like Gallium-Arsenic and Silicon substrates have also been used. [100,101] In a molecular or more precisely single molecule electronic device, the other components in the circuit also need to be molecular in nature. Therefore, molecular wires, which are generally made of conjugated polymers, are used to facilitate the charge transfer across the device. Interestingly, among the molecular wires category, alkane-based and polyphenylacetylene (OPE)-based wires have been the most widely studied systems.[102] However, two-dimensional OPE wires suffer from intermolecular π-π stacking problems.[103] Such interactions make conductance measurement for single wire difficult as it may lead to the attachment of multiple molecules within the molecular junction. In this context, ferrocene-based molecular wires having three-dimensional ferrocene unit are advantageous over OPE wires.[104]

Crivillers and co-workers have successfully incorporated ferrocene-derivative in molecular electronic junctions to investigate the photoisomerization of a styrene-based system in solution and on the surface.[105] Inclusion of ferrocene unit helps in preventing photo-activated generation of phenanthrene and also provides platform for electrochemical characterization. They demonstrated that *trans* to *cis* transformation becomes much more efficient when molecule was attached on Au elctrode. The group isolated pure *trans* and *cis* forms of the molecule and prepared SAMs which were then independently studied in the form of EGaIn/Ga$_2$O$_3$/*trans*(or **cis**)ferrocene-SAM/Au (**Fig. 26a-b**). Current-voltage measurements showed about 1.3 times higher current density for *cis*-SAMs over *trans*-SAMs (**Fig. 26c**). Observed difference in current densities might arise due to difference in molecular length and the dipole moments. In another set of experiments, isomerically pure *cis*- and *trans* -Fc SAMs were investigated by illuminating with suitable light. *Trans*-Fc-SAMs were irradiated using 340 nm wavelength light, and current density was measured before and after illuminating for 40 minutes (**Fig. 27**). It was observed that before irradiating the sample, maximum value (log|J|≈ 2.1) corresponds to pure *trans*-Fc-SAMs, whereas, after irradiation the maximum value shifted to log|J|≈ 1.2 which corresponds to *cis*-Fc-SAMs with a *trans*- to *cis*- conversion was calculated to be around

82%. This study represents that electrochemical investigation could be applied to study photo-isomeric SAMs on the surface.

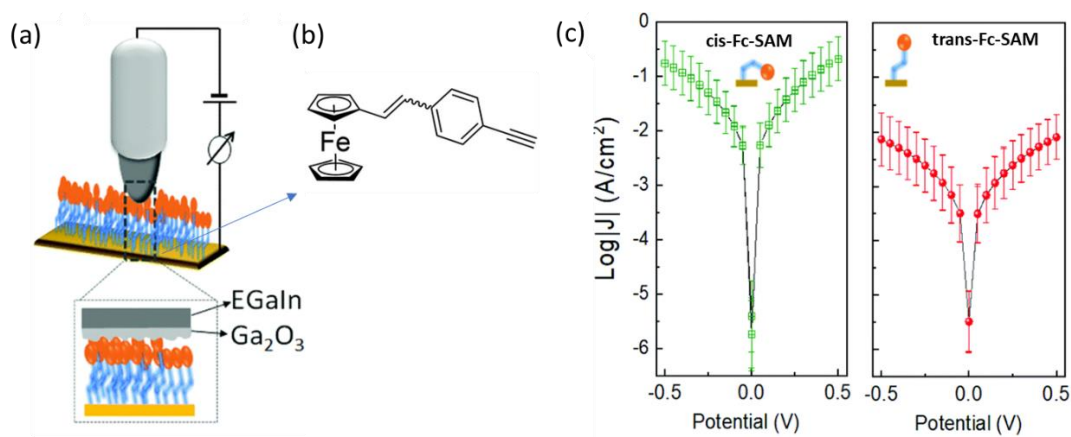

**Fig. 26.** (a) Schematic diagram showing a ferrocene-SAMs MJs, (b) chemical structure of ferrocene containing molecule, (c) logJ-V plot for *cis-* and *trans-*Fc-SAMs MJs. Reproduced with permission from Ref. [105]Copyright 2022, Royal Society of Chemistry.

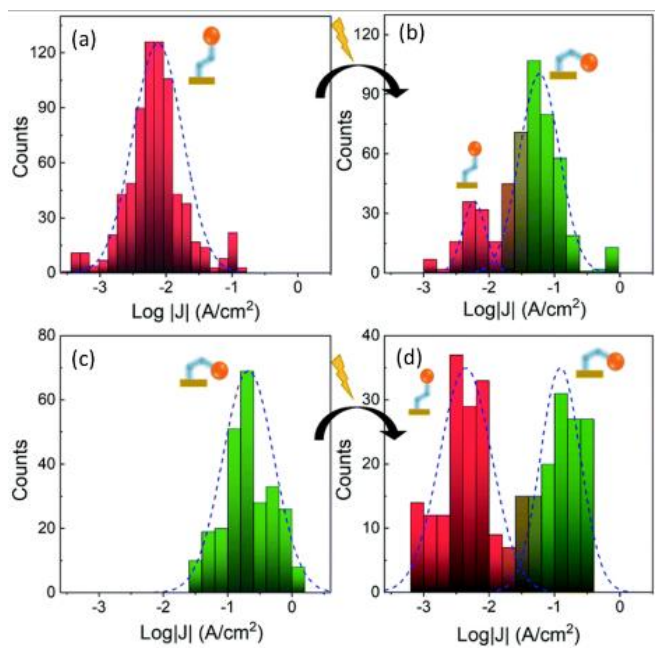

**Fig. 27.** (a-b) *trans-* to *cis-* conversion of Fc-SAMs (λ=340 nm, 40 minutes) and current density distribution, (c-d) *cis-* to *trans-* conversion following irradiation for 40 minutes with a 254 nm wavelength light. Reproduced with permission from Ref. [105] Copyright 2022, Royal Society of Chemistry.

Zhang and co-workers incorporated ferrocene into an organic skeleton an prepared an organometallic small molecule-based rewritable memory device (**Fig. 28**).[106] Ferrocene attached to a pyrene containing conjugated molecular platform (Py-Fc) exhibited a bistable switching operation with long retention time. The sweep-mode I-V measurements confirmed the FLASH binary propertied of the device. The switching time of the fabricated device was determined to be 20 μs and even after $10^3$ cycles, no signs of fatigue was observed. The device reproducibility was confirmed by testing fifty units that further confirmed the suitability of Py-Fc based device to be used for binary flash storage purposes.

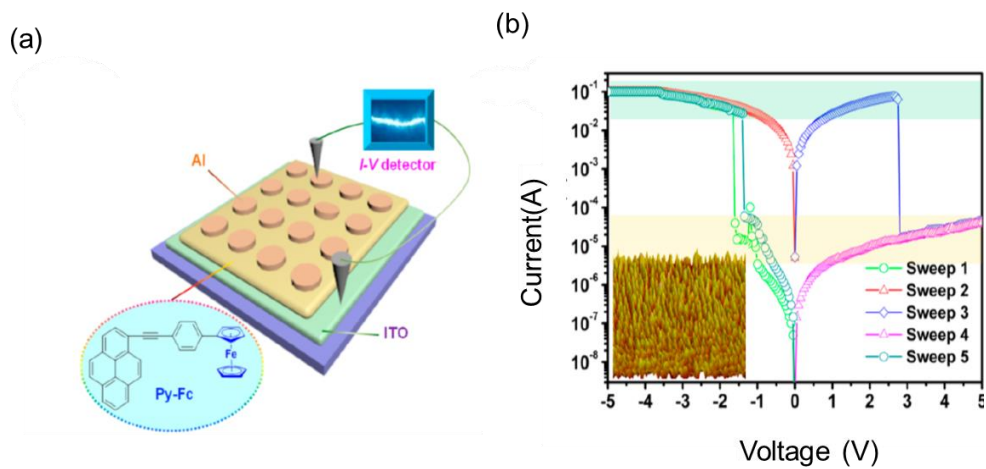

**Fig. 28.** (a) Depiction of the MOM-type OSMD structure, (b) I-V characterisations of the OSMDs. Reproduced with permission from Ref.[106], Copyright 2019, American Chemical Society.

Grutter and co-workers used ferrocene (Fc) moiety attached to hexadecanethiol to describe nuclear quantization in electron transfer between single molecule and gold substrate by experimental and theoretical methods.[107] Electron transfer was measured using oscillating AFM technique keeping tip to molecule distance of about 10 nm to avoid any electron tunnelling. A popular way of constructing SAM-based molecular junctions is to seal the SAMs between two ultra-flat electrodes at top and bottom. This design makes the exploitation of intrinsic molecular property of SAMs inaccessible due to the presence of two bulk electrodes causing isolation of the SAMs from external environment. Duan and co-workers prepared a ferrocene-based SAMs tunnel junction using Au and single layer graphene electrodes (**Fig. 29a**).[108] The vertical tunnel junction thus created allowed in situ control over redox properties of ferrocene unit. Redox reactions of ferrocene groups present beneath the graphene layer could be controlled by applying suitable oxidant, reductant or electrochemical stimuli at the top. In case of oxidation of ferrocene units, the counter anions at graphene surface act to balance the charge. **Fig. 29b** shows the current density ($J_d$) Vs source-drain voltage ($V_d$) plots before and after addition of oxidizing agents ($H_2O_2$) and then after addition of reducing agent ($NaBH_4$). Addition of $H_2O_2$ causes curve to become more symmetrical and significant reduction in $J_d$ values. However, upon addition of reducing agent, the curve restores its shape to becoming almost identical to initial stage. This could be reversibly switched between oxidized and reduced states.

Nijhuis and co-workers investigated the ferrocene/EGaIn-based molecular tunnel junctions to unravel information about energy level alignment. [109] In order to understand the nature of charge transport whether it is resonant tunnelling or off-resonant type, charge transport across this junction was measured using normalized differential conductance (NDC). In past few decades, inclusion of DNA in molecular electronics has attracted much attention. Xu and co-workers found that a molecular rectifier obtained by intercalating coralyne in to a 11-base pair DNA duplex exhibited a rectifier ratio of about 15. [110] The basic assumption that DNA exists in stretched form in the device, made in most of the previous studies, have caused various disagreements amongst the research community. In order to avoid such issues, related with order-disorder of the DNA, Nijhuis and co-workers investigated the charge transport properties in a disordered monolayers containing single-stranded DNA (ssDNA) and an ordered monolayers containing double stranded DNA (dsDNA).[111] Electrical properties of these DNA monolayers supported over Au-electrode and EGaIn electrode (**Fig. 30**). They prepared DNA containing 15, 20, 25 and 30 base pairs to have four kinds of samples with varying sequence lengths. The Au-linker-DNAn-Fc//GaOx/EGaIn junctions were then subjected to J(V) measurement to yield influence of the number of base pairs in the system (**Fig. 30c-d**). Further analysis of the obtained data revealed absence of any specific trend with increasing number of pairs in case of ssDNA, whereas, same fitting of the data for dsDNA-based junction showed an exponential decay. This discrepancy is attributed to the flexibility in ssDNA structure leading to folded structures and therefore length dependency remains unclear.

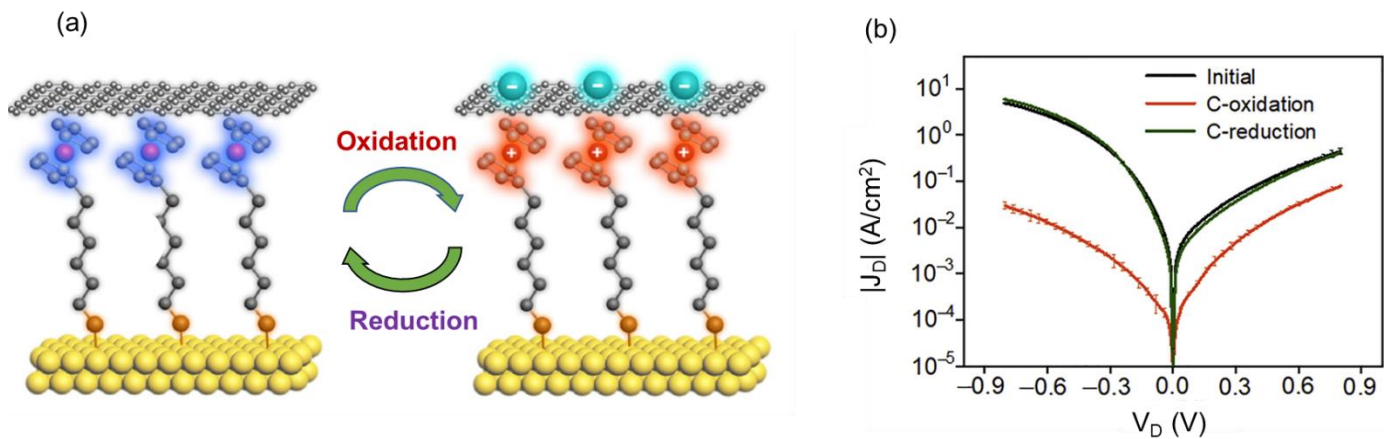

**Fig. 29.** (a) Schematic depiction of Au/Fc-SAM/SLG junctions along with oxidation and reduction events. (b) current density vs source-drain voltage for the device upon treatment with both oxidising and reducing solutions. Reproduced with permission from Ref. [108], Copyright 2020, Elsevier.

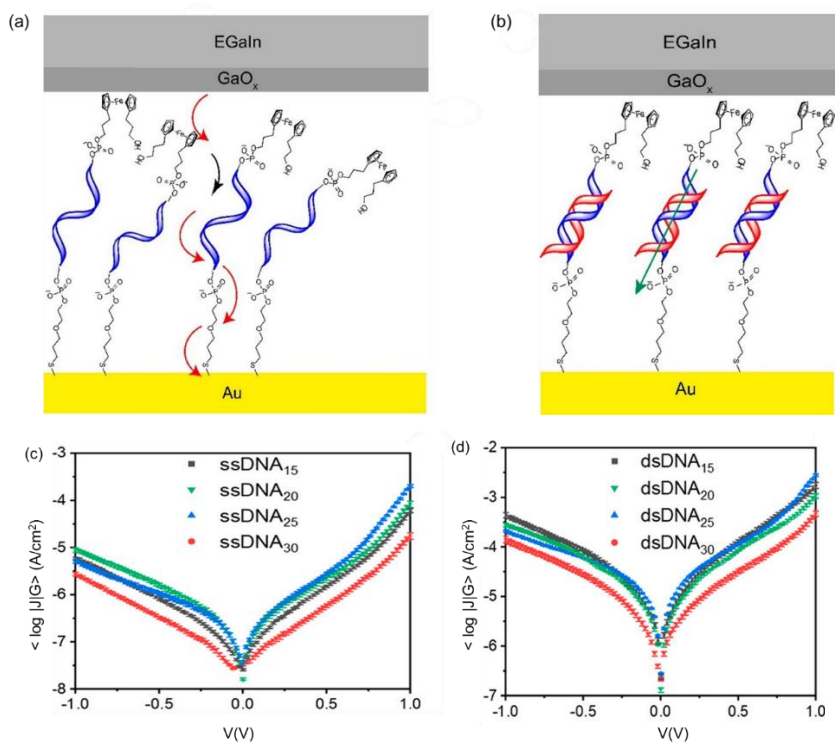

**Fig. 30**. (a) Au-ssDNA$_{15}$-Fc//GaOx/EGaIn junction showing intra (red) and inter (black) hopping using curvy arrows. (b) Au-dsDNA$_{15}$-Fc//GaOx/EGaIn junction showing coherent tunnelling utilising a green arrow. (c-d) corresponding logJ-V plots. Reproduced with permission from Ref. [111], Copyright 2021, American Chemical Society.

# 8. Conclusions and Future Perspective

Ferrocene remained a structural puzzle for some time after its discovery in 1951. It took no time to realize the capabilities of this newly found aromatic system and indeed ferrocene was extensively studied for its catalytic activity along with its redox properties. Ferrocene being the first sandwiched-type structure, pioneered a new era of organometallic complexes known as "metallocenes". Some of the ferrocene's unique features such great stability in ambient conditions, benign nature and ease of functionalization made it to be used as a molecular scaffold to create new molecules for very diverse applications ranging from materials science to biomedical chemistry. In materials chemistry ferrocene is used a standard in electrochemical measurements. The excellent reversible redox propertied exhibited by ferrocene have made it as one of the most widely used building block in stimuli-responsive systems. Ferrocene combined with optically active systems results in interesting photo-electric effects. Redox properties offered by ferrocene has also been used in preparing dye-sensitized solar cells and a large number of molecular electronics-based devices. In recent past, ferrocene has been studied for its medicinal properties as well. The range of applications and variety of materials that could be obtained using ferrocene is vast. It is very correct to say that ferrocene as a redox marker holds huge potentials in designing exciting new materials for future applications in devices and biomedicine. After almost seventy-five years of its discovery, research activity around ferrocene is still very active as can be seen from huge number of reports published every year. Ferrocene is a potent molecular building block and a prospective structural unit.

**Acknowledgment**

GR would like to thank Ashoka University for her financial support and research funding. RG thanks IIT Kanpur for providing a Senior Research Fellowship. SRS would like to thank the University Grants Commission of India (UGC) for providing a Junior Research Fellowship. SS wishes to thank Prof. S. Basu, Director, CSIR-Institute of Minerals & Materials Technology, Bhubaneswar, India, for in-house financial support (Grant number: CSIR-IMMT-OLP-112) and requisite permissions. PCM acknowledges the Department of Science and Technology for a start-up research grant (SRG/2019/000391), and IIT Kanpur (IITK/CHM/2019044) for initiation and special grant support to establish the laboratory. Financial support by the Council of Scientific & Industrial Research (CSIR), New Delhi, Sanctioned NO.:01(3049)/21/EMR-II awarded to PCM is highly acknowledged. The authors wish to thank Dr. Priyajit Jash, a former post-doc with PCM group for designing TOC artwork.